\documentclass[sigchi]{acmart}

\usepackage{booktabs} % For formal tables
\usepackage{multirow}
\usepackage{hyperref}

\setcopyright{rightsretained}
\acmDOI{.}

\acmISBN{.}

\acmConference[Under Review]{}{}{}

\acmYear{2019}
\copyrightyear{2019}

\begin{document}
\title{ClinicalVis: Supporting Clinical Task-Focused Design Evaluation} % in the intensive care unit}
\subtitle{A Visualization-Based Prototype System to Explore the Interaction of Healthcare Providers and Electronic Healthcare Records}
%\titlenote{Produces the permission block, and  copyright information}
%\subtitlenote{The full version of the author's guide is available as \texttt{acmart.pdf} document}

% %% This is how authors are specified in the journal style

\author{Marzyeh Ghassemi}
\affiliation{%
  \institution{Verily, University of Toronto}
  \city{Toronto}
  \state{Ontario}
}
\email{marzyeh@cs.toronto.edu}

\author{Mahima Pushkarna}
\affiliation{%
  \institution{Google Brain}
  \city{Cambridge}
  \state{Massachusetts}
}
\email{mahima@google.com}

\author{James Wexler}
\affiliation{%
  \institution{Google Brain}
  \city{Cambridge}
  \state{Massachusetts}
}
\email{jwexler@google.com}

\author{Jesse Johnson}
\affiliation{%
  \institution{Verily, Sanofi}
  \city{Cambridge}
  \state{Massachusetts}
}
\email{}

\author{Paul Varghese}
\affiliation{%
  \institution{Verily}
  \city{Cambridge}
  \state{Massachusetts}
}
\email{paulvarghese@verily.com}

\renewcommand{\shortauthors}{M. Ghassemi et al.}

\hypersetup{
    colorlinks=true,
    linkcolor=blue,
    filecolor=magenta,      
    urlcolor=cyan,
}
 
\urlstyle{same}

\begin{abstract}
Making decisions about what clinical tasks to prepare for is multi-factored, and especially challenging in intensive care environments where resources must be balanced with patient needs. Electronic health records (EHRs) are a rich data source, but are task-agnostic and can be difficult to use as summarizations of patient needs for a specific task, such as ``could this patient need a ventilator tomorrow?'' In this paper, we introduce \textit{ClinicalVis}, an open-source EHR visualization-based prototype system for task-focused design evaluation of interactions between healthcare providers (HCPs) and EHRs. We situate ClinicalVis in a task-focused proof-of-concept design study targeting these interactions with real patient data. We conduct an empirical study of 14 HCPs, and discuss our findings on usability, accuracy, preference, and confidence in treatment decisions. We also present design implications that our findings suggest for future EHR interfaces, the presentation of clinical data for task-based planning, and evaluating task-focused HCP/EHR interactions in practice.
\end{abstract}

% %
% % The code below should be generated by the tool at
% % http://dl.acm.org/ccs.cfm
% % Please copy and paste the code instead of the example below.
% % \ccsdesc[500]{Computer systems organization~Embedded systems}
% % \ccsdesc[300]{Computer systems organization~Redundancy}
% % \ccsdesc{Computer systems organization~Robotics}
% % \ccsdesc[100]{Networks~Network reliability}


\maketitle

\section{Introduction}
Patient Electronic Health Records (EHR) contain a wealth of heterogeneous data, leading to exciting opportunities in both information visualization and clinical support. Planning for safe and effective clinical care depends on the ability to parse this heterogeneous clinical data, and derive an understanding of a patient's health state. Data visualization techniques are known to improve rich data pattern communication and reduce overall cognitive load~\cite{shen2002data}, which in turn can help Health Care Providers (HCPs) efficiently extract accurate information~\cite{card1999readings}. However, commercially available EHRs are often cognitively cumbersome to use, and EHR usability is a well-established HCP pain point~\cite{doi:10.1001/jama.2018.1171}. 

Commercial EHRs are task-agnostic, support linear (rather than dynamic) care coordination processes, and are optimized for billing rather than HCP and patient use~\cite{o2010electronic,bates2010getting}. They are also closed-source, making them difficult to evaluate~\cite{khairatassessing}. Even taking screen images of a leading EHR provider can be viewed as legally prohibited~\cite{Kenyon1,Tahir1}. Access to clinical data --- and clinical environments --- is often limited for privacy reasons, leading professionals tasked with improving EHR designs to work with secondary information rather than real-world use cases~\cite{khairatassessing}. 

There is currently no open-source framework that provides a visual summary of patient information to HCPs for planning specific clinical task, and subsequently evaluates HCP response. To support the evaluation of HCPs-EHR interaction for task-specific clinical planning, our multidisciplinary team assessed a need to (a) select realistic clinical tasks HPCs need to plan for, (b) create a prototype system to view EHRs on, (c) identify a set of real patient cases where received care was recorded for each task, and (d) characterize HCP-EHR interaction during care planning in both the prototype and a baseline system. 

\begin{figure*}[t]
  \centering
  \includegraphics[width=\linewidth]{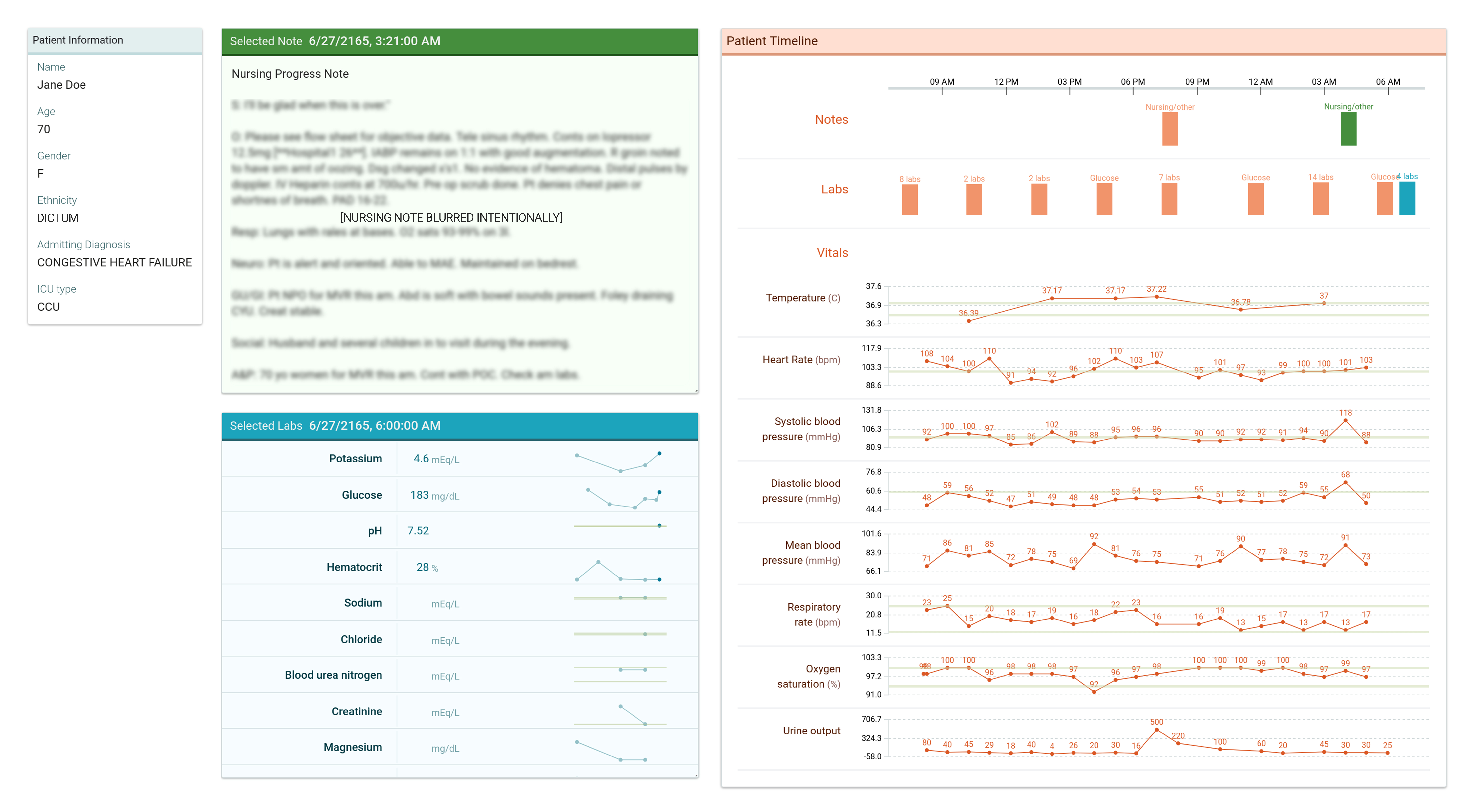}
   \caption{A screenshot of the \textit{ClinicalVis} user interface.  Nursing note intentionally obscured to protect privacy. Actual de-identified patient records were displayed during the experiments.}
   \label{fig:modelOverview}
\end{figure*}

In this work, we present \textit{ClinicalVis} (Figure~\ref{fig:modelOverview})\footnote{\url{http://github.com/PAIR-code/clinical-vis }}, an open-source, and freely available visualization-based prototype system with a proof-of-concept design study~\cite{sedlmair2012design} that we validated with empirical case studies of practicing HCPs and real patient data. ClinicalVis was designed by our team for the purposes of understanding and supporting task-focused interactions between HCPs and EHRs. We compare ClinicalVis to a baseline system, designed to emulate commercially available EHR interfaces that HCPs used in practice at the time of the study. The design study focuses on the use of EHRs for clinical decision making by HCPs for meaningful clinical tasks. Specifically, we use real, anonymized patient EHRs from the MIMIC-III dataset~\cite{johnson2016mimiciii} in the two interfaces to observe the HCP-EHR interactions while focused on a realistic clinical scenario --- a remote HCP asked to plan care for physiological decompensation amongst multiple patients in an intensive care unit (ICU). 

In this paper, we outline related work, present the prototype systems (ClinicalVis and the baseline system), and describe methods used to evaluate HCP interaction with the interfaces. Our findings and insights from are briefly as follows. First, we found that HCP accuracy in forecasting patient needs from EHR was generally poor, and information overload was not overcome by an improved task-agnostic visualization. Second, ClinicalVis improved HCP experience during the tasks, and in a post-task comparison. Finally, we noted that HCPs using ClinicalVis spent their time validating care plans rather than finding information in the EHR, and present considerations for how future work may augment in-situ HCP care planning. 
%TODO: UPDATE ONCE MAHIMA UPDATES THE PAPER!

In summary, the contributions of this paper are:
\begin{itemize}
\item Providing an open-source EHR visualization-based prototype system for task-focused design evaluation at \url{http://github.com/PAIR-code/clinical-vis}. 
\item Summarizing the findings of a proof-of-concept design study targeting HCP-EHR interactions in a task-focused setting, conducted with real patient data on practicing HCPs. 
\item Describing insights for future EHR interfaces in presenting clinical data for care planning, and evaluating their efficacy in practice. 
\end{itemize}

%     First, we found that HCP accuracy in forecasting patient needs from EHR was generally poor, and information overload was not overcome by an improved taskagnostic visualization. Second, ClinicalVis improved HCP experience during the tasks, and in a post-task comparison. Finally, we found that HCPs using ClinicalVis spent their time validating care decisions rather than finding information in the EHR, and present considerations for how future work may augment in-situ HCP clinical decision making

% I'd suggest making these your bullet point list of contributions; the current bullet points are at a high level and probably don't communicate your contributions as well as the points listed above.

\section{Related Work}
We summarize past work on care planning complexity, HCP information needs, evaluation methods for task-focused care planning, and information visualization techniques in EHRs. 

\subsection{Care Planning Complexity in the ICU}
Care planning in the ICU is challenging; clinical signals are often irregularly sampled and contaminated by interference and human error. Information visualization and chart reviews become particularly meaningful for decision making in an environment like the ICU where clinicians must process patient information quickly \cite{blum2010alarms}, and there is near-constant information from multi-modal devices, and multi-disciplinary staff \cite{sutcliffe2004communication}. Data complexity, and the difficulty of working with EHRs has introduced new threats to patient safety~\cite{han2005unexpected}, including the introduction of error from fragmented displays and alert fatigue~\cite{national2012health}. Prior studies have demonstrated that 80\% of ``user error'' (correlated to 12-22\% annual mortality rate) can be attributed to cognitive overload~\cite{rothschild2005critical,henriksen2008understanding}. In this work, we evaluate our prototype system in a realistic task setting to understand the influence of visual EHR summaries in clinical care planning.

\subsection{Supporting HCP's Information Needs} 
As the quantity of useful electronic health data burgeons, the ability to efficiently review, interpret, and assimilate diverse data types becomes critical to planning patient care\cite{mazur2016toward}. Current EHR designs contribute to common errors and information loss via mistaken patient identification, mode mismatches, flawed interpretations, incorrect recall or incomplete visibility of system states \cite{shneiderman2013improving}. Further, EHR usability is poor~\cite{doi:10.1001/jama.2018.1171}, and contributing factors are not well-understood~\cite{ellsworth2016appraisal}. EHR software builds in hospitals are often the result of multiple information systems intersecting, with information generated from multiple sources, such as physicians, nurses, billing staff, and even by the patients themselves \cite{Hyrinen2008DefinitionSC}. Inadequately designed bedside interfaces are the greatest contributor to diagnostic cognitive load~\cite{sinsky2012comparative,chou2012health}, and the ability to customize information display for different users and varying clinical scenarios is crucial~\cite{nolan2017multisite} due to convoluted workflows and prolonged searching activities~\cite{nolan2017health}. 

\subsection{Task Based Evaluation of Care Planning}
Clinical Decision Support (CDS) systems have been suggested for care planning, but such systems have faced several barriers in adoption. \citet{middleton2016clinical} suggest that CDS dissatisfaction is caused by challenges in aligning an HCP's mental model of the patient, diagnostic process, and therapeutic care plans. Well-designed systems that support decision making steps could bolster CDS integration into HCP's everyday workflows~\cite{rajkomar2011improving}. Evaluating HCP-EHR interaction is often with Task Load Index (TLX) \cite{hart1988development} measures to study the workloads of clinical tasks, and investigate the cognitive demand of EHR transitions \cite{colligan2015cognitive}. TLX measures have previously been used to study mental, physical and cognitive workload in 17 clinical participants performing a set of clinical tasks on simulated patients in three clinical diagnosis scenarios: urinary tract infection, pneumonia and heart failure \cite{mazur2016toward}. The authors' results suggest that task demands as experienced by HCPs (e.g., needing more clicks, requiring more time) are related to performance (e.g., more severe omission errors) regardless of EHR type. Follow-up work with the same tasks, and quantifying effort towards task and omission errors strengthened these results \cite{mosaly2018relating}. In our paper, we specifically target simple prototype designs to focus study on the impact of HCP-EHR interactions during task-based care planning.

\subsection{Visualization-based Explorations of EHR Design} 
Information visualization techniques are known to help people carry out tasks more efficiently by highlighting aspects of the data \cite{card1999readings} that might require pattern recognition \cite{savikhin2008applied}, as well as reducing cognitive load and freeing up working memory in decision making \cite{cook2005illuminating}. Visual summaries that allow users to explore and analyze their data have a rich and successful standing in infectious disease epidemiology, but these tools are often deeply siloed for specialized applications \citet{carroll2014visualization}. There is also a strong body of prior work on designing visualization-based systems and prototypes that support of teamwork in healthcare coordination \cite{amir2015care}, and that support patients in managing their own care \cite{klasnja2010blowing,ballegaard2008healthcare}. There are also many commercial EHR visualizations, such as EPIC and Apple's Health Dashboard. Such work is promising, but duplicating the evaluations conducted on closed-source systems are difficult, and experiments on care planning in a task-oriented setting are hard to reproduce. 

There is comparatively little research addressing EHR visualization in a realistic task-focused scenario. Within care planning, implementation of the AWARE EHR visualization system saved time on patient chart review \cite{pickering2015implementation} and was associated with improved patient outcomes in the ICU \cite{olchanski2017can}, demonstrating that streamlined interfaces can improve the efficiency and reliability of clinical staff \cite{ahmed2011effect}. Likewise, systems like LifeLine~\cite{plaisant1996lifelines}, Timeline~\cite{bui2007timeline} and MIVA 2.0~\cite{faiola2015}, have demonstrated the power of visualizing clinical information visualization using a common time scale for multi-modal clinical data streams. Other systems dynamically scale time intervals instead so that data scales may be modified by user interaction, e.g., VISITORS~\cite{klimov2010intelligent} and Midgaard~\cite{bade2004connecting}. In these cases, systems were evaluated as they are, but not within the focus of a simulated clinical task on real patient data. We differ from these systems in that we focus on simple time-constrained visualizations without any prompts or notifications, to conduct task-focus evaluations of how clinicians move through their workflow through our interface.  

\section{Design of Prototype Systems}
We designed and implemented ClinicalVis (Figure~\ref{fig:modelOverview}), a visualization-based prototype system that supports evaluation of HCP-EHR interaction during realistic clinical care planning tasks using iterative and participatory design methods. Here, we discuss the process and final design of our prototype system.
%MAR, can you get the last sentence reviewed by fanny/josef?

\subsection{ClinicalVis Design Process} 
ClinicalVis is designed as a content-independent but structure-dependent system to enable rapid information assimilation, and support the inference of insights from large amounts of data. We ground our design decisions in known and observed HCP workflows ~\cite{nolan2017health}, workplace-specific emergent practices  ~\cite{zhou2009just,heath1996documents}, and basic requirements outlined in \citet{heath1996documents}. We further targeted our designs to mitigate common challenges faced by HCPs in using visual summaries and EHRs~\cite{carroll2014visualization,roman2017navigation}. We iteratively stress-tested the system through development, and internally validated our designs against the expertise of our multi-disciplinary team, which includes a practicing physician, machine learning with EHRs expert, and data visualization and interaction design experts. 

\subsection{ClinicalVis Interface and Interactions}
ClinicalVis enables HCPs to explore the most recent 24 hours of patient information in four visually-distinct modules, to arrive at a diagnostic care plan for a displayed patient case. The layout and scale of the modules are determined by the physician's reading order, and interaction capabilities are limited to low-level interactions such as scrolling, clicking and hovering. Further, modules can be re-sized for HCP comfort using drag-and-drop. %(Figure~\ref{fig:clinicalvisLabsHover}, Figure~\ref{fig:clinicalvisVitalsHover}) 

Clinical notes, ordered labs, and observed vital signs are marked at the time of entry on a 24-hour timeline in a large, single module. Vitals are visualized as a line chart with datapoints at the time of observation, and lab orders and notes are marked as bars at the time of log. The most recent clinical/imaging note and lab is loaded by default in the note- and lab- specific modules, and users can load different notes or labs into appropriate modules by clicking on the 24-hour timeline. The X-axis denotes time moving from left to right, and individual Y-axes denote the max and min for vitals values. Individual labs are visualized as sparklines inside the lab module.  Missing or unavailable lab values are not visualized. Although all visualized datapoints for labs and vitals have labels, we included hovering capabilities to enable HCPs to individually read a datapoint's timestamp and numeric value.

Information is color coded by type (vital/lab/note/patient identity) and demographic-agnostic guidelines \cite{horowitz2010defining} indicate abnormal values when applicable. Color is sparsely used for encoding at higher levels of abstraction within the modules, and primarily reserved for conveying associations or relaying feedback between different information across modules.

\subsection{Baseline GUI Design}
To understand the influence of ClinicalVis in care planning vis-a-vis commercially available EHRs, we created an imitation control graphical user interface (GUI) as our baseline with the same information and interaction capabilities as ClinicalVis. We model the design of the Baseline GUI (Figure~\ref{fig:baselineOverview}) after in-use EHR systems that several authors have first-hand experience with, and knowledge of. 

We minimize the influence of any out-of-scope elements by constraining both interfaces to (a) be self-contained on a single page with no external links, (b) support a single patient record at a given time and (c) display a curated set of patient vitals pre-reviewed for task-specific evaluation. During evaluation, HCPs were shown real and unedited clinical data from patient cases in both interfaces. To limit the scope of this study, we exclude any spatial or imaging data (such as X-rays or ultrasound scans), however retaining any associated textual or free-form data.

\section{Research Setting and Methods}
Here we describe the study design, data and task choices, participant cohort, and study methods. 

We formulated our evaluation to study two complementary research questions: 
\begin{enumerate}
\item How do HCPs engage with visual representations of real patient EHR data during task-focused care planning?
\item Do visual representations of EHRs influence clinical care planning --- specifically, do they impact accuracy, confidence and time-to-decision?
\end{enumerate}

\subsection{Data Source}
Clinical data are fundamentally multi-modal, and many different data types are relevant to understanding patient health \cite{weber2014finding}. We use data from the Medical Information Mart for Intensive Care III (MIMIC-III) database \cite{johnson2016mimiciii}. MIMIC-III is publicly available, and contains over 58,000 hospital admissions from approximately 38,600 adults. Prior visualization work on the open-source MIMIC dataset have focused on web-based tools for researchers, e.g. an interface to identify cohorts for study creation \cite{lee2015web}, and predictive mortality modules \cite{chen2015explicu,levy2018visualizing}. Our paper is a novel use of MIMIC-III records to evaluate HPC interaction with EHR systems, and the impact that has on care planning. 

\subsection{Task Definition} 
We emulated an eICU\footnote{The ``eICU'' is a clinical term for an ``electronic'' or remote ICU that a supporting HCP is making judgments from.} setting, where multidisciplinary teams of HCPs must forecast care needs to prepare on-site HCPs for therapeutic interventions~\cite{celi2001eicu}. We focused on common tasks for physiological decompensation that have potential risks: mechanical ventilation for breathing assistance~\cite{yang1991prospective,tobin2006principles}, and vasopressor administration to regulate a patient's blood flow~\cite{mullner2004vasopressors,d2015blood}. 

In our evaluation, we used records of ICU patients aged 15 and older that met two criteria: (a) The patient record did not display any target interventions in the ICU for at least 24 hours before the 8 AM start of ``rounds''\footnote{Morning rounds are a group evaluation of patient needs and care planning in the ICU.}, and for at least 12 hours after rounds, \emph{and} (b) Each patient record had at least 1 note during the 24 hours prior to the rounds. %All records were aligned by absolute (clock) time to emulate an ICU environment.  

From this subset, we selected 1 EHR for training, 2 EHRs for the think aloud, and 8 positive and 8 control patients \footnote{Demographic details of the selected patients are shown in Table~\ref{table:SelectedPatients}.} of equal estimated difficulty for the proof-of-concept study as follows:
\begin{itemize}
\item \textbf{VE+}: First ventilation 4 - 12 hours after rounds. 
\item \textbf{VP+}: First vasopressor 4 - 12 hours after rounds.
\item \textbf{Control (C)}: No ventilation or vasopressor in the 12 hours after rounds.
\end{itemize}

\subsection{Participant Cohort}
We recruited 14 clinicians practicing in hospitals in a large metropolitan area who submitted valid responses to a recruitment form. Each potential participant was pre-screened for prior ICU experience (averaging 9 hours per week in the ICU) and the cohort was controlled for diversity of specialization. 
%ICU experience amongst participants ranges from experience gained during rotations to specializing in critical care. Participants generally tend to spend approximately 9 hours on average per week in an ICU setting. 
A summary of our participants' self-reported demographics is available in Table~\ref{table:ParticipantExperience}.
All participants were invited to a laboratory setting for evaluation during the timeframe of the study and successfully completed the study in entirety.

\begin{table}[!ht]
\label{tab: experience}
\centering
  \begin{tabular}{llll}
 \toprule
    {\small \textit {ID}}
        &{\small \textit {Specialization}}
        &{\small \textit {Experience}}        
    &{\small \textit {ICU Time}} \\
    & & {\small \textit {Years}} & {\small \textit {Hours/Week}} 
    \\
    \midrule
    {\small {P1}} & N/A & 1+ & < 4  \\
    {\small {P2}} & Pediatrics & 2+ & < 4 \\
    {\small {P3}} & Infectious diseases & 2+ & 20 to 24 \\
    {\small {P4}} & General Surgery & 3+ & 4 to 8  \\
    {\small {P5}} & Critical Care & 4+ & 16 to 20  \\
    {\small {P6}} & Hospital Medicine & 4+ & < 4 \\
    {\small {P7}} & Emergency Medicine & 4+ & 8 to 12 \\
    {\small {P8}} & Internal Medicine & 5+ & 4 to 8 \\
    {\small {P9}} & Pediatric Critical Care & 5+ & 16 to 20 \\
    {\small {P10}} & Cardiology & 5+ & 16 to 20 \\
    {\small {P11}} & Cardiology & 5+ & < 4 \\
    {\small {P12}} & General Medicine & 7+  & < 4 \\
    {\small {P13}} & General Practitioner & 10+ & < 4  \\
    {\small {P14}} & Critical Care & 10+  & 20 to 24  \\
 \bottomrule
\end{tabular}
\label{tab:experience}
\caption{Participant breakdown by self-reported specialization, years in current role, and number of hours per week spent in the ICU.}
  \label{table:ParticipantExperience}
\end{table}
% \begin{figure}[h!]
%   \centering
%   \includegraphics[width=0.75\linewidth]{chart__1_.png}
%   \caption{The self-reported experience levels of participant cohort.}
%   \label{fig:docs}
% \end{figure}
% A first pilot study was conducted on 3 clinicians to ensure clarity of scenario prompt and instructions, following which the study was conducted on 14 pre-screened participants. The final study was split into three sections. Participants were introduced to the scenario prompt and task. Then, in the first section, participants were asked to perform a think-aloud protocol as they familiarized themselves with both GUIs. 

\subsection{Study Design}
\begin{figure*}[!ht]
  \centering
  \includegraphics[width=0.95\linewidth]{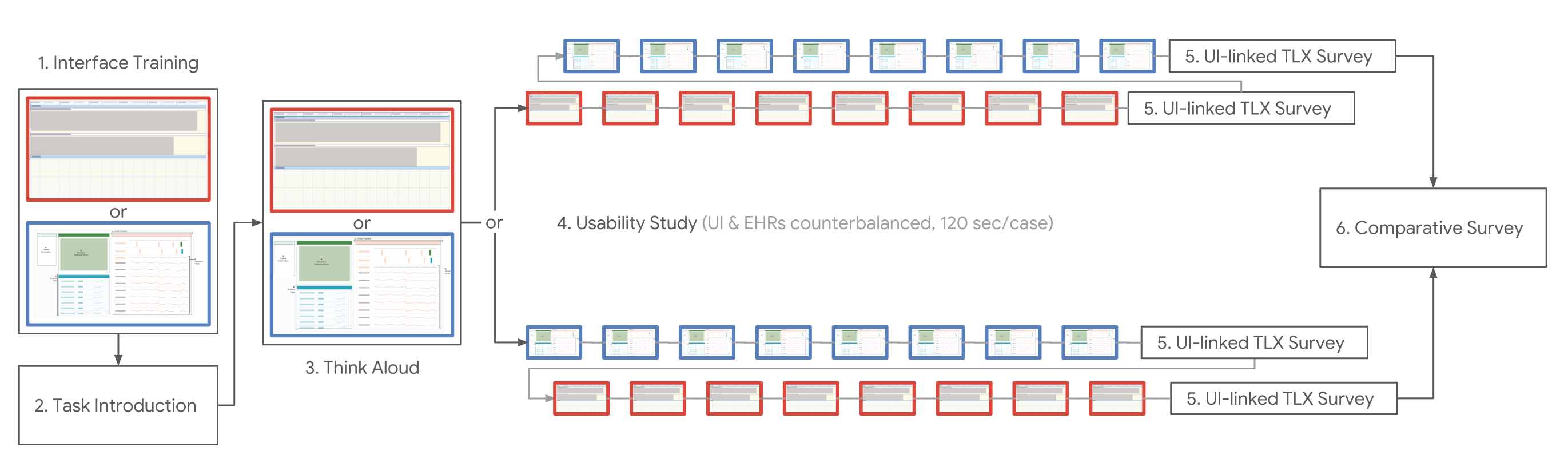}
  \caption{Experimental flow. Red denotes the Baseline prototype, blue denotes ClinicalVis. Participants were walked through a scripted introduction (1 \& 2) to the study, which was divided into three broad sections: A \textit{think-aloud} protocol (3), a task-focused usability study (4) \& TLX surveys (5), and a post-completion comparative feedback survey (6). Participants were assigned randomly generated alphanumeric codes as identities to capture data and feedback consistently across the study.}
  \label{fig:expFlow}
\end{figure*}

We conducted an empirical evaluation of HCP-EHR interaction with ClinicalVis and the baseline prototypes in a mixed-methods, task-focused user study centered around clinical care planning a remote eICU scenario. An overall experimental flow in shown in Figure~\ref{fig:expFlow}.  

Participant sessions lasted approximately 50 minutes, and were conducted in a lab setup at (\textit{redacted for review}). After introducing participants to the study, both prototypes, and the task (5 mins), we conducted a think aloud to qualitatively evaluate the differences in interactions with EHRs between ClinicalVis and the baseline (10 mins). We then conducted a usability study (35 mins) and TLX survey on both interfaces to track how ClinicalVis supported HCPs in care planning. Finally, we used an open-ended comparative survey to capture participant experience (10 minutes).  

\subsubsection{1. Training}
At the start of each session, participants were introduced to the goals of the study, the sources of the data, and notified that the data was real de-identified patient information that had not been modified or cleaned. After outlining the three-part structure of the study. participants were introduced to the baseline and ClinicalVis in counterbalanced order on the same patient record. A high-level explanation of the layout and interaction capabilities of both interfaces was provided. 

\subsubsection{2. Task Introduction}
To reduce incentives for providing positively-skewed feedback, we asked participants to perform realistic care-planning in a time-sensitive scenario (possible physical decompensation in each patient record) using the following prompt:

{\small \textit{``You are a clinician in an eICU and have just come on-shift. You manage decompensation alerts for two hospitals, each with its own EHR. You are remotely shown 24 hours of the available vitals, labs and notes, and cannot request more. You are asked to review the records of patients from each hospital. Records from Hospital A will look different than those from Hospital B due to the EHR variance. For each patient, you will decide if staff should be prepared for a Vasopressor, Ventilator, both, or neither in the next 12 hours, and indicate how confident you are.''}}

\subsubsection{3. Think aloud}
We conducted \textit{think-alouds} to gain qualitative insights into HCP interactions with the baseline and ClinicalVis prototypes, and the usability challenges faced in planning care using both interfaces. \cite{lewis1993task,holzinger2005usability}. Each participant was shown one pre-selected EHR in counterbalanced order per prototype (also counterbalanced). For each prototype, they were provided with a prompt to think aloud, and a list of assisting questions (see Appendix \ref{sec:ThinkQuestions}). No time limit was enforced, and participants were free to clarify questions about the prototypes, task or patient record.

\begin{figure}[h]
  \centering  
  \includegraphics[width=0.8\linewidth]{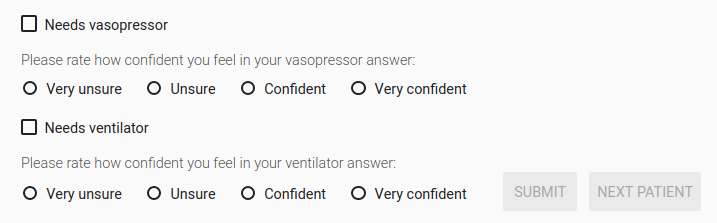}
  \caption{GUI for the evaluation after viewing each patient.}
  \label{fig:question}
\end{figure}

\subsubsection{4. Usability Evaluation and 5. TLX Survey}
%Each UI shows 24 hours of patient data for 8 patient EHRs for 120 seconds in counterbalanced order with additional time allowed in increments. 
Participating HCPs were assigned to review cases for each prototype sequentially, in counterbalanced order. The ordering of 8 patient records per interface was also counterbalanced using Latin Squares\cite{keedwell2015latin}. HCPs were presented with ICU data from the preceding 24 hours of each patient record, and had 120 seconds\footnote{HCPs often plan care through patient-centered reflective actions \cite{fischer2001communities}, and therefore limit the time that each participant has with a patient record.} to arrive at a decision. Resets were permitted in increments of 120 seconds. HCPs were asked to submit a \textit{''yes''} or \textit{''no'} decision for each intervention (vasopressor and ventilator), and indicate their confidence in each decision on a 4-point likert scale with no neutral choice in an evaluation screen \ref{fig:question} at the 2-minute mark. HCPs could open the evaluation screen before the end of the time limit, but could not return to the same case without a reset.

Upon the completion of 8 consecutive patient assessments, participants were asked to complete a survey for the prototype they interacted with. The survey comprised Likert scales measuring mental demand, physical demand, effort level, temporal demand, performance and frustration as defined by the NASA Task Load Index \cite{hart1988development} on a 10-point scale.

\subsubsection{6. Comparative Survey}
Task assessments can lack interpretation, even when observed directly \cite{dix2009human}. To capture retrospective feedback after having interacted with both prototypes, participants were asked to compare ClinicalVis and the Baseline along the TLX axes, and optionally provide a reason for their selection. Additionally, participants asked to indicate which of the two prototypes supported the task better. Rationales were solicited as an open-ended free-text responses with no word limit.

\subsubsection{Measures}
To evaluate these research questions, we collected the following measures:
\begin{itemize}
\item \textbf{Accuracy (\%):} Rate of correct responses for a given case for a participant.
\item \textbf{Time to Task (seconds):} Time taken to arrive at a decision for a given case. This includes any resets.
\item \textbf{Confidence:} Self-reported confidence in a decision for a given case using a 4-point Likert scale, scored between -2 (not at all confident) and 2 (very confident).
\item \textbf{TLX scores:} Self-reported mental demand, physical demand, effort level, hurriedness, success and discouragement on a scale of 0-10.
\end{itemize}

\begin{table*}[!t]
  \centering
  \begin{tabular}{rrlllll}
 % \toprule
  & & \multicolumn{3}{c}{By Group}
    & \multicolumn{2}{c}{Across All}\\
     \cmidrule(lr){3-5}\cmidrule(lr){6-7}
  & & {\smaller \textit{Vasopressor}} & {\smaller \textit{Ventilator}} & {\smaller \textit{Control}} & {\smaller \textit{Vasopressor}} & {\smaller \textit{Ventilator}} \\
  &  & {\smaller \textit{Needed}} & {\smaller \textit{Needed}} & {\smaller \textit{(None)}} & {\smaller \textit{Correct}} & {\smaller \textit{Correct}} \\
%\midrule
\cmidrule(lr){3-5}\cmidrule(lr){6-7}
\multirow{ 2}{*}{Accuracy (\%)}
  & {\small{Baseline}} & 50.00\% & 56.25\% & \textbf{71.64\%} & 62.50\% & 55.35\% \\
  & {\small{ClinicalVis}} & \textbf{68.83\%} & \textbf{62.79\%} & 67.64\% & \textbf{63.30\%} & \textbf{58.92\%} \\
\cmidrule(lr){3-5}\cmidrule(lr){6-7}
\multirow{ 2}{*}{Confidence}
  & {\small{Baseline}} & 0.68 & 0.87 & 1.34 & 1.14 & 0.98 \\
  & {\small{ClinicalVis}} & \textbf{1.41} & \textbf{1.27} & \textbf{1.47} & \textbf{1.28} & \textbf{1.09}\\
\cmidrule(lr){3-5}\cmidrule(lr){6-7}
\multirow{ 2}{*}{Avg. Time to Task (s)}
  & {\small{Baseline}} & 92.31s & 92.73s & \textbf{83.64s} & \multicolumn{2}{c}{87.11s per case/HCP} \\%TODO: merge last two cells
  & {\small{ClinicalVis}} & \textbf{84.43s} & \textbf{86.86s} & 85.37s & \multicolumn{2}{c}{\textbf{85.94s} per case/HCP} \\
  \cmidrule(lr){3-5}\cmidrule(lr){6-7}

 %\bottomrule
  \end{tabular}
  \caption{Accuracy, confidence and time-to-task results of average participant performance per case using the Baseline and ClinicalVis prototypes. The best results for each comparison are bolded.}
  \label{table:TaskStatistics}
\end{table*}

%From initial discussions with HCPs, we focused on variables were most likely to be used when predicting vasopressor or ventilator requirements and ultimately extracted 5 static variables (e.g., gender), 30 time-varying vitals and labs (e.g., oxygen saturation), and all available, de-identified clinical notes for each patient as a timeseries across their entire stay. See Section \ref{sec:variables} in the Appendix for a complete listing.
%We utilized grounded theory as a basis to conduct the analysis of all responses but do not report on the process in detail for space reasons. 
% >>>
% ``The presentation of the qualitative results was anecdotal and unclear.  I have no
%     problem with the inclusion of example quotes, but the proliferation of qualifiers
%     like "often" or "several participants" left me unclear as to the strength of any
%     trends.  I completely failed to understand the comments about the blood pressure
%     labeling at the end of section 6.1.2.
% ''
% Accuracy, usability, preference and confidence. Reporting time 2 task, accuracy, confidence levels, and TLX results (preference).
% * <mahimapushkarna@gmail.com> 2018-09-20T19:48:30.733Z:
% 
% why 65%? what is a good accuracy?
% 
% ^.

\section{Results and Discussion}
In this section we discuss our findings and elaborate on how those suggest design implications for future work on EHRs and HCP-EHR interaction, discussing these results in turn. Table~\ref{table:TaskStatistics} summarizes our quantitative results from 224 simulated EHR-HCP encounters (112 for each prototype). 

Our key insights are primarily related to practice, specifically that (a) EHR to task accuracy in care planning is generally poor, and, counter-intuitively, did not improve with a better visualization; (b) visualizations changed the way HCPs experience data; and (c) HCPs maximized time for validation when interacting with visual summaries.

\subsection{EHR-To-Task Performance is Generally Poor}
Our hypothesis was that an improved visual interface would lead to better HCP performance on planning care for realistic tasks, where better performance was characterized by increased accuracy and/or lowered time-to-decision. Our analysis revealed that overall performance was better on ClinicalVis than on the Baseline, but insignificantly so.

\begin{figure}[hpb]
  \centering
  \includegraphics[width=0.8\linewidth]{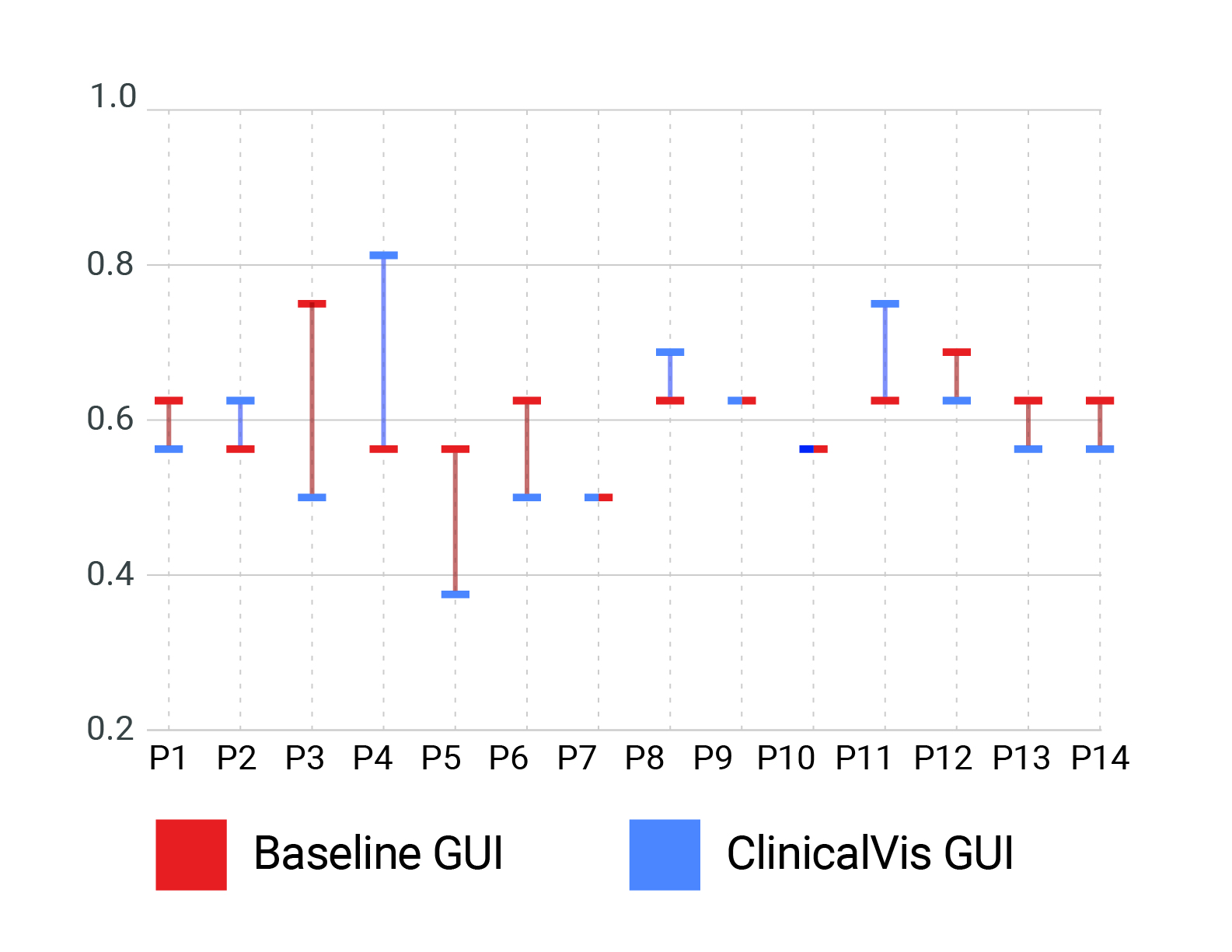}
  \caption{Individual HCP accuracy in care planned with Baseline and ClinicalVis, higher is better.}
  \label{fig:accuracy}
\end{figure}

Participant's average accuracy was higher when using ClinicalVis (63.30\%)  compared to the baseline (62.50\%), and overall accuracy remained under 75\%. Further, there was no observable pattern in accuracy across individual participant performance for the two interfaces (Fig\ref{fig:accuracy}).

Though self-reported HCP confidence in planned care was significantly higher when using ClinicalVis (1.18 vs 1.06 for baseline), the average time-to-decision was insignificantly lower in ClinicalVis (85.94s/case/HCP) than the baseline (87.11/case/HCP). 

\subsubsection{EHR Are Fundamentally Limited For Care Planning}
While prior work has found clinical and economic benefits in adopting eICUs~\cite{coustasse2014business}, many of the signals factored into care planning come from being able to physically interact with the patient~\cite{celi2001eicu}.

"\textit{Part of the frustration with EHRs in general, is that people try and make predictions based on the data, which is just not as helpful as laying eyes on the patient}" - P4

Seven participants echoed that not seeing the patient reduced their confidence considerably and this influenced the amount of caution exercised, with one participant estimating that ``\textit{90\% of signals come from physically observing the patient}''.  Further, in the absence of the patient and/or data, seven participants reported that they were either re-reading, over-thinking or reading too much into the admitting diagnosis, EHR or prognosis.

"\textit{eICUs are actually of questionable effectiveness, because there is really only so much info that you can get from the data. With any one of these examples, my confidence would go up considerably if I actually saw the patient.}" - P4

While our work addresses the need to integrate well within an HCP's workflow, our findings show that even when evaluating individual cases, an HCP's information needs are complex and constantly changing. The physical absence of the patient, unavailability of data and inconsistencies within the data were cited as key causes for frustrations and delays in interacting with EHRs on both interfaces.

\subsubsection{Fundamental Data Concerns that affect HCPs}
Participant feedback indicated that the key factors influencing clinical preparedness in our evaluation were rooted in the underlying EHR data. Many EHRs have \textit{incomplete} or \textit{inconsistent} data in the underlying source record, reflecting actual available information~\cite{ghassemi2018opportunities}. Sources for these include different intervals for aggregation, discrepancies between the time of an observation and time of logging the observation into the system, and hospital- or ICU- specific practices.

"\textit{(It is) The nature of data that prevents me from making a decision, such as lack of knowledge of interventions.}" - P5

"\textit{I look at what's going on in the last 4 hours because nurses won't put everything at the same time though they try.}" - P1

HCPs interacted with prototype systems with reduced confidence in cases \textit{they perceived} as having insufficient data. Inconsistencies in data presentation and non-standardized data logging practices further exacerbated frustrations and reduced confidence during interactions with an EHR for both interfaces. For instance, five participants requested that urine output be as an aggregate rather than in absolute units - however, each indicated different preferences for the time interval (1 hour, 12 hours, 24 hours) for aggregation. Additionally, two participants informed us that they were unfamiliar with abbreviations in the notes. One participant observed that \textit{"notes written in all uppercase make me believe that the nurse is yelling at me."}

\subsection{Better Visualization Improved HCP Experience}
Participants reported that ClinicalVis had little impact on their overall performance, but they experienced the data differently. In line with our design goals, we found that participants preferred ClinicalVis over the Baseline. They reported lower cognitive loads, reduced opportunities for error, increased content engagement and information retrieval. 

\begin{figure}[ht]
  \centering
  \includegraphics[width=1\linewidth]{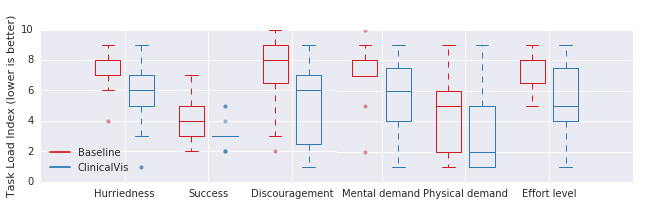}
  \caption{Individual responses to each prototype for the TLX survey. Lower scores are better, and ClinicalVis (blue) consistently had improved responses.}
  \label{fig:tlxBoxplots}
\end{figure}
\subsubsection{TLX analysis}
A stronger preference for ClinicalVis was evident on all TLX dimensions (Figure~\ref{fig:tlxBoxplots}). We computed the the Kolmogorov-Smirnov statistic \cite{massey1951kolmogorov} on the two samples (ClinicalVis and baseline) to test for significance in each question. We found that ClinicalVis passed statistical significance ($p=0.10$) level for feelings of discouragement ($p=0.017$) and effort ($p=0.051$). However, at this level, participant preference for ClinicalVis on mental demand, physical demand, hurriedness, and success was not statistically significant. 

Prior work has found that EHR-related task demands significantly increases mental effort and task difficulty, which are predictors of omission error performance \cite{mosaly2018relating}. One participant remarked that\textit{ "[I] felt reluctant reading the table [baseline] and going into the minor details... I was more comfortable making decisions having engaged with the data at a greater level [in the visualization]."} (P13). %While overall performance was low in our evaluation, 
Increased loads reported by the HCPs for the baseline can be viewed as a proxy for the number of general omission errors likely to be made by systems in practice. 

\subsubsection{Experiencing EHRs differently}
We observed that ClinicalVis integrated well into physician work flow, and physicians reported being able to establish clearer clinical narratives, find responses to questions faster, and identify outliers and relevance with lower effort. Participants were able to quickly habituate themselves to our prototype system. 
Interactions were learned quickly, and all 14 participants reported spending less attention on the interface and more on the data. In contrast, nine participants described their performance on the last few patient records in the baseline as poor due to "mental fatigue", "excessive effort" or "exhaustion". One participant stated that \textit{"the baseline felt very demanding for patients that were not very sick"}, while also noting that \textit{"This one (CinicalVis) felt slower, I came to a decision faster so I tried to slow myself down and I think I was over-thinking it, but it was very clear if the patient was going to need a vasopressor or a ventilator or if they were going to decompensate."}

\subsubsection{Data-first interface}
\begin{figure}[ht]
  \centering
  \includegraphics[width=0.9\linewidth]{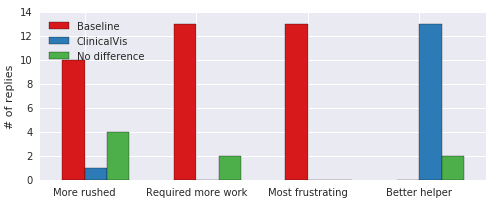}
  \caption{Post-study comparative evaluation of the Baseline and ClinicalVis on TLX dimensions. Note that in the first three comparisons of \textit{negative} affect the Baseline scores higher. In the final comparison of \textit{positive} affect, ClinicalVis scores highest.}
  \label{fig:tlxBars}
\end{figure}

Given the distributed nature of tasks in an ICU, HCPs frequently rely on interpersonal trust within and between the ICU team and consulting services to plan care~\cite{zhang2002designing}. In the absence of such support, HCPs felt varying degrees of frustration, effort and assistance with both prototypes (Fig~\ref{fig:tlxBars}). %However, task analysis can be used to uncover opportunities for improving usability as users self-solve problems \cite{kuniavsky2003observing}.  
HCPs preferred ClinicalVis to the baseline across four different metrics: 1) which visualization made the task feel more rushed, 2) which required more work in completing the task, 3) which was most frustrating to use, and 4) which was a better at supporting the task. Two participants felt no difference in effort across the two interfaces, instead attributing the effort applied to the patient cases.

\subsubsection{Altered perception of content }
During the comparative evaluation, participants tended to focus on the increased mental demand required from the baseline, noting that they felt rushed and anxious about finding relevant information:  \textit{``Certain sense of anxiety... trying to make these predictions is hard, and when you feel like you're having to fight with the UI to find what you are looking for.'}. When asked about factors determining the indicated confidence levels in the baseline, the general sentiment pointed towards a fear of having missed a number, or a lack of confidence in the data itself. Such concerns are well-established in prior work on HCP-EHR interaction, in which a majority of surveyed clinicians were worried about overlooking important information due to the volume of data and inadequate display/organization \cite{nolan2017health}. In contrast, the ClinicalVis system allowed them to have more confidence in their ultimate assessments of the patient.\textit{"Some interfaces in hospital are (pause) unfortunately like the baseline and they make you feel like you have no control. It makes you nervous. The visualization is much more reassuring."} 

Participants' comments about the baseline were primarily centered around finding and perceiving information, whereas feedback on ClinicalVis tended to describe the status of the patient, indicating deeper engagement with the data.  
%Engagement with patient data in ClinicalVis increased HCP confidence during care planning. 
Participants had a tendency to use terms such as ``\textit{spikes}'', ``\textit{dips}'', ``\textit{trending up}'' to describe the patient data rather than describing the visualization, suggesting that information was being consumed at a faster rate. Further, all participants claimed that notes was easier to read and four participants assumed that it had been extrinsically modified to improve reading, though they had not. 

\subsubsection{Clinical Confidence in Visualization}
Prior work has indicated that HCPs primarily focused on the viewing of clinical notes during electronic chart reviews (ECR) in the ICU (44.4\% of their time), followed by laboratories (13.3\%), imaging studies (11.7\%), and searching/scrolling (9.4\%) for typical cases \cite{nolan2017health}. 47\% of ECRs also began with review of clinical notes, which were the most common navigation destination.

Participant's interactions with the data in our interface echoed these findings, with two exceptions. First, all physicians started with reviewing patient information. We can speculate that additional attention was paid to this module due to the influence of the study and the absence of the physical patient. %However, since this behavior persisted for all ClinicalVis tasks, we believe that errors attributed to incorrect patient identification can be mitigated in ClinicalVis. 
Secondly, the review of clinical notes was heavily interspersed with the viewing of vitals and labs. In particular, physicians looked to clinical notes to iteratively answer questions that arose from the charts, and moved quickly between the patient timeline module and the notes module. Two participants were observed contextualizing parts of the nursing note within the patient timeline, explaining that notes were typically created over the course of a few hours and logged at a later time. In such scenarios, we find that a single-screen and modular approach to information presentation creates tighter feedback cycles, enabling  HCPs to confirmed their intuitions with comments in the notes. \textit{``The graphic interface is much more helpful, the separation of subjective/ objective, and the trending function is just better in the visualization.''} When asked about effort, participants emphasized the difficulty of locating information in the baseline. \textit{"It's harder to find the information that you are looking for, harder to see trends and separate out the different components that you are looking at."}

\begin{figure}[ht!]
  \centering
  \includegraphics[width=1\linewidth]{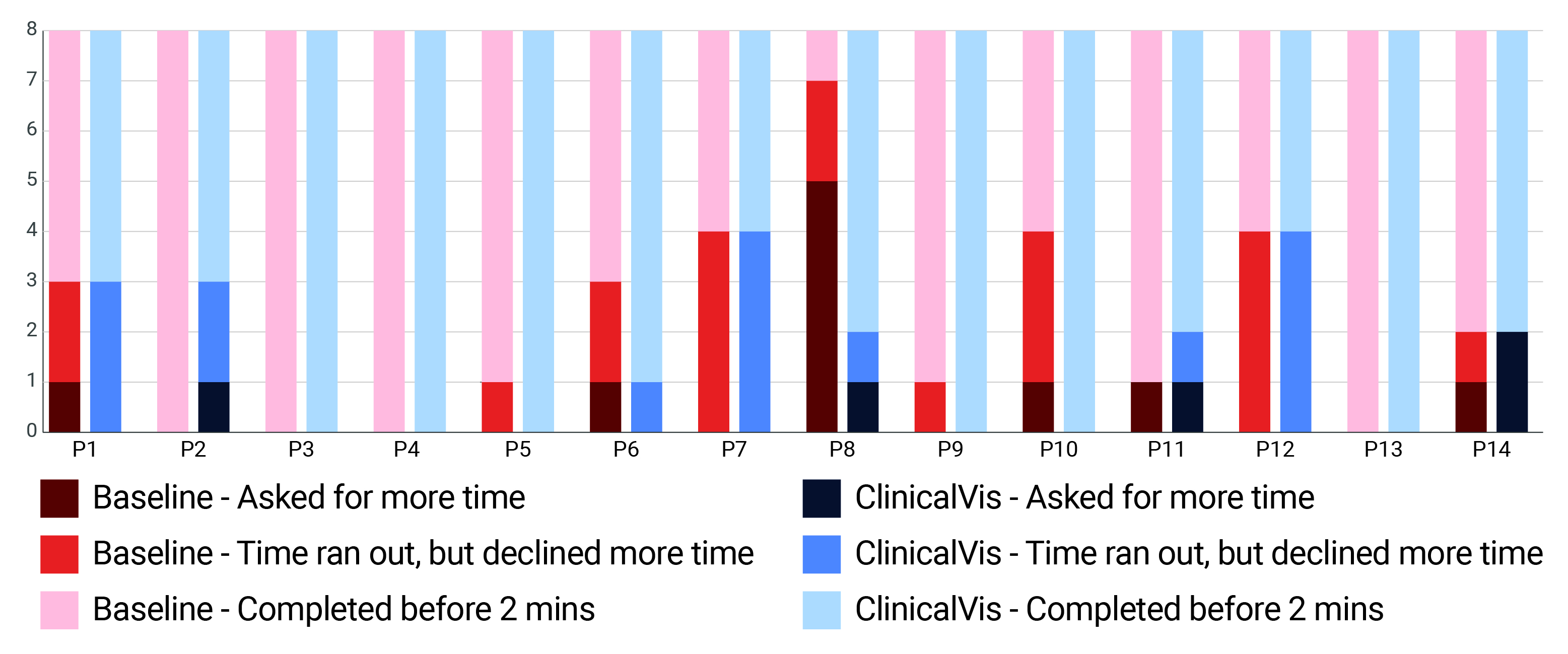}
  \caption{Time requirements per participant for all tasks in the Baseline compared to ClinicalVis. Note that HCPs often asked for more time (dark red) when using the Baseline.}
  \label{fig:baselineCount}
\end{figure}

\subsection{Maximizing Time For Decision Validation, Not Information Finding}
In time-sensitive environments, the efficient use of time is paramount to care planning. Maximizing the efficiency of time and attention spent with each patient record in ClinicalVis allowed participants to verify facts contributing to care plans. 

\subsubsection{Inadequacies of Time to Task metric}
HCPs spent time in \emph{validating their decisions} in ClinicalVis, compared to time spent in \emph{looking for information} in the baseline (Fig \ref{fig:baselineCount}).  \textit{``Because I spent more time looking... so I had less time with the table.'}.  We found that ClinicalVis gave participants a sense that they were able to digest data more completely, and time remaining after having arrived at a decision was used for validation as opposed to submitting in the evaluation screen. In the light of this behavior, the insignificant difference between the average time to task per case per HCP between the two interfaces suggests that time to task was an inadequate measure of performance.

\subsubsection{Influence of individual assessments of caution}
We anticipated that HCPs care about treating the patient, through patient-centered reflective actions \cite{fischer2001communities}, and therefore may increase cautious planning. However, prior work has been unable to prove that routine reflection in action can improve diagnostic outcomes\cite{norman2009dual}. In our study, we find that while providing avenues in EHR interfaces that support active reading and routine reflection did not increase outcome accuracy, the increased time-to-task and reduced accuracy for control cases suggests a higher false positive rate. Participant descriptions attribute this to increased levels of caution and clinical preparedness for ``a prognosis that could go either way". 

It was observed that participating physicians exercised varying levels of caution in planning care when unsure, with some being more conservative than others. Some physicians were more cautious and willing to prepare resources regardless of confidence in their decision. For instance, in one case, a participant noted that they felt unsure of prognosis, even though the patient appeared stable --- the participant chose to prepare for vasopressor administration. For the same case, another participant was "\textit{confident how this (prognosis) would go}" and did not prepare for either, correctly.

\subsubsection{Validation practices that support heuristic methods}
Prior work has indicated that HCPs apply a hypothetico-deductive method for arriving at a diagnosis during a patient encounter \cite{elstein1978medical}, in which diagnostic hypotheses are ``proposed, tested, and either verified or rejected''. This is further augmented by various "mental shortcuts" or heuristics that can sometimes lead to diagnostic errors. \cite{rajkomar2011improving}. We observed several of these heuristics in practice when using ClinicalVis and the baseline. Particularly, HCPs frequently exhibited the use of the representative heuristic\footnote{Used to estimate the likelihood of a condition based on how closely a patient's presentation matches prototypical cases of the condition. \cite{rajkomar2011improving}} in both interfaces. However, the related heuristics of anchoring \footnote{A tendency to be attached to initial diagnostic hypothesis despite the accumulation of contradictory evidence.}, premature closure\footnote{Settling on a diagnosis with insufficient investigation of information.} and confirmation bias\footnote{Seeking evidence that supports working hypotheses, ignoring contradictory or ambiguous evidence.} appeared to be mitigated through data validation practices. 

HCPs were frequently observed to verbally change  their decisions in ClinicalVis as they traversed different parts of the visualization; further, they demonstrated better recall of patient information in the evaluation screen; finally, twelve participants noted that it was easier to establish clinical narrative using ClinicalVis. The validation experience is a critical because reflection and communication are parts of clinical experience that technology currently hinders, and prior work has demonstrated that 91\% of all medical errors can be attributed to difficulties in communication and collaboration \cite{cohen2006cognitive}.

\section{Limitations and Future Work}
Our work is a first step towards an evaluation of visualization systems that focuses on the task-oriented setting in which clinicians work, and the environmental factors they face. Here we detail limitations and the future work that is needed. 

First, our work limited the displayed data to the most direct indicators for the chosen care planning tasks. A more robust exploration of data sources and types should be integrated into the ClinicalVis system, accounting for time-scale and sampling differences. Second, we created ClinicalVis with simple design principles, without an extensive design-focused process. Future work is needed to investigate whether visual representations that mimic denotations and reading practices of HCPs would improve the evaluated measures and outcomes. Third, we investigated two specific tasks in a specific care planning setting. More complete sets of decision making situations should be tested --- both in terms of the tasks and scenarios. Fourth, our sample of HCPs was limited, and we believe that a larger sample size of HCPs should be examined to validate our findings for a broader population. Finally, our finding that accuracy did not improve with better visualization of patient data was interesting and unexpected, and should be studied further.

\section{Conclusion}
In this paper, we present an empirical evaluation of ClinicalVis, a visualization-based prototype system, on the interactions of HCPs with EHRs in a task-focused setting. We then present insights learned from 14 participating HCPs as they interacted with EHRs in a simulated eICU to plan for care for real patient cases using ClinicalVis. Specifically, we found that (a) counter-intuitively, EHR-to-task was generally poor in the physical absence of patients, (b) ClinicalVis positively altered the way HCPs experienced data without significantly impacting performance ,and (c) physicians preferred to maximize available time by purposing it for decision-validation in ClinicalVis. Importantly, HCPs using ClinicalVis reported increased confidence, which is particularly relevant as the field of machine learning begins to target the use of EHR data to answer specific clinical questions~\cite{wu2010prediction,ghassemi2018opportunities,steele2018machine}. 

%Our study consisted of think-alouds, usability studies and TLX and comparative surveys. 
%In this paper, we present the insights learned from HCPs as they triage real patient cases using ClinicalVis and a baseline GUI emulating present-day EHRs. 
%We chose a visual summary within a task-focused scenario because 
% First, we establish that HCPs are able to use both a baseline system and \textit{ClinicalVis} to make accurate and confident care plans when presented with data from real ICU patients. Second, we find that \textit{ClinicalVis} streamlines information processing, and provide qualitative supporting evidence. Finally, we demonstrate that HCPs prefer \textit{ClinicalVis} to the baseline system---rating it higher both in individual rankings during decision making, and competitive rankings post-task. %ClinicalVis led to improved HCP experience, but did not significantly improve task accuracy.  
%Importantly the system streamlined information processing, allowing HCPs to focus on validating their decision. 

While we chose data that was most appropriate for predicting physiological decompensation, our proof-of-concept interface is agnostic to the specific experimental task performed. ClinicalVis is scalable to include variables other than the ones presented; our visualization is not custom-fit for this task, and can be used more generally to investigate during and post-task clinical usage of ICU EHR. The design implications of our work suggest that modeling clinical data for decision support should include elements to guide clinical use, and to that end, we have open-sourced ClinicalVis as a means to support and engender such efforts. We believe that ClinicalVis can open up avenues for the rigorous evaluation of interactions between clinicians and patient data to best improve aspects of healthcare delivery.

%Our multidisciplinary team of clinicians, researchers, and bioinformatics experts are excited to design and test new tools that can help, rather than hinder, clinical care. 

% This is a key opportunity for visualizations in the ICU to be transformative, and create different ways of thinking about EHR. As eICUs and remote support become more population, using visualizations to augment signals data and provide a better sense of patient state are valuable contributions. 

\begin{acks}
We would like to thank several people who contributed to making this work substantially better, including Lauren Dukes, Michael Terry, and Joseph Jay Williams.
\end{acks}

\bibliographystyle{ACM-Reference-Format}
\bibliography{clinical_vis}


\begin{thebibliography}{70}

%%% the Web address.

\ifx \showCODEN    \undefined \def \showCODEN     #1{\unskip}     \fi
\ifx \showDOI      \undefined \def \showDOI       #1{#1}\fi
\ifx \showISBNx    \undefined \def \showISBNx     #1{\unskip}     \fi
\ifx \showISBNxiii \undefined \def \showISBNxiii  #1{\unskip}     \fi
\ifx \showISSN     \undefined \def \showISSN      #1{\unskip}     \fi
\ifx \showLCCN     \undefined \def \showLCCN      #1{\unskip}     \fi
\ifx \shownote     \undefined \def \shownote      #1{#1}          \fi
\ifx \showarticletitle \undefined \def \showarticletitle #1{#1}   \fi
\ifx \showURL      \undefined \def \showURL       {\relax}        \fi
% invisible to TeX
\providecommand\bibfield[2]{#2}
\providecommand\bibinfo[2]{#2}
\providecommand\natexlab[1]{#1}
\providecommand\showeprint[2][]{arXiv:#2}

\bibitem[\protect\citeauthoryear{Ahmed, Chandra, Herasevich, Gajic, and
  Pickering}{Ahmed et~al\mbox{.}}{2011}]%
        {ahmed2011effect}
\bibfield{author}{\bibinfo{person}{Adil Ahmed}, \bibinfo{person}{Subhash
  Chandra}, \bibinfo{person}{Vitaly Herasevich}, \bibinfo{person}{Ognjen
  Gajic}, {and} \bibinfo{person}{Brian~W Pickering}.}
  \bibinfo{year}{2011}\natexlab{}.
\newblock \showarticletitle{The effect of two different electronic health
  record user interfaces on intensive care provider task load, errors of
  cognition, and performance}.
\newblock \bibinfo{journal}{\emph{Critical care medicine}}
  \bibinfo{volume}{39}, \bibinfo{number}{7} (\bibinfo{year}{2011}),
  \bibinfo{pages}{1626--1634}.
\newblock


\bibitem[\protect\citeauthoryear{Amir, Grosz, Gajos, Swenson, and Sanders}{Amir
  et~al\mbox{.}}{2015}]%
        {amir2015care}
\bibfield{author}{\bibinfo{person}{Ofra Amir}, \bibinfo{person}{Barbara~J
  Grosz}, \bibinfo{person}{Krzysztof~Z Gajos}, \bibinfo{person}{Sonja~M
  Swenson}, {and} \bibinfo{person}{Lee~M Sanders}.}
  \bibinfo{year}{2015}\natexlab{}.
\newblock \showarticletitle{From care plans to care coordination: Opportunities
  for computer support of teamwork in complex healthcare}. In
  \bibinfo{booktitle}{\emph{Proceedings of the 33rd Annual ACM Conference on
  Human Factors in Computing Systems}}. ACM, \bibinfo{pages}{1419--1428}.
\newblock


\bibitem[\protect\citeauthoryear{Bade, Schlechtweg, and Miksch}{Bade
  et~al\mbox{.}}{2004}]%
        {bade2004connecting}
\bibfield{author}{\bibinfo{person}{Ragnar Bade}, \bibinfo{person}{Stefan
  Schlechtweg}, {and} \bibinfo{person}{Silvia Miksch}.}
  \bibinfo{year}{2004}\natexlab{}.
\newblock \showarticletitle{Connecting time-oriented data and information to a
  coherent interactive visualization}. In \bibinfo{booktitle}{\emph{Proceedings
  of the SIGCHI conference on Human factors in computing systems}}. ACM,
  \bibinfo{pages}{105--112}.
\newblock


\bibitem[\protect\citeauthoryear{Ballegaard, Hansen, and Kyng}{Ballegaard
  et~al\mbox{.}}{2008}]%
        {ballegaard2008healthcare}
\bibfield{author}{\bibinfo{person}{Stinne~Aal{\o}kke Ballegaard},
  \bibinfo{person}{Thomas~Riisgaard Hansen}, {and} \bibinfo{person}{Morten
  Kyng}.} \bibinfo{year}{2008}\natexlab{}.
\newblock \showarticletitle{Healthcare in everyday life: designing healthcare
  services for daily life}. In \bibinfo{booktitle}{\emph{Proceedings of the
  SIGCHI Conference on Human Factors in Computing Systems}}. ACM,
  \bibinfo{pages}{1807--1816}.
\newblock


\bibitem[\protect\citeauthoryear{Bates}{Bates}{2010}]%
        {bates2010getting}
\bibfield{author}{\bibinfo{person}{David~W Bates}.}
  \bibinfo{year}{2010}\natexlab{}.
\newblock \bibinfo{title}{Getting in step: electronic health records and their
  role in care coordination}.
\newblock
\newblock


\bibitem[\protect\citeauthoryear{Blum and Tremper}{Blum and Tremper}{2010}]%
        {blum2010alarms}
\bibfield{author}{\bibinfo{person}{James~M Blum} {and} \bibinfo{person}{Kevin~K
  Tremper}.} \bibinfo{year}{2010}\natexlab{}.
\newblock \showarticletitle{Alarms in the intensive care unit: too much of a
  good thing is dangerous: is it time to add some intelligence to alarms?}
\newblock \bibinfo{journal}{\emph{Critical care medicine}}
  \bibinfo{volume}{38}, \bibinfo{number}{2} (\bibinfo{year}{2010}),
  \bibinfo{pages}{702--703}.
\newblock


\bibitem[\protect\citeauthoryear{Bui, Aberle, and Kangarloo}{Bui
  et~al\mbox{.}}{2007}]%
        {bui2007timeline}
\bibfield{author}{\bibinfo{person}{Alex~AT Bui}, \bibinfo{person}{Denise~R
  Aberle}, {and} \bibinfo{person}{Hooshang Kangarloo}.}
  \bibinfo{year}{2007}\natexlab{}.
\newblock \showarticletitle{TimeLine: visualizing integrated patient records}.
\newblock \bibinfo{journal}{\emph{IEEE Transactions on Information Technology
  in Biomedicine}} \bibinfo{volume}{11}, \bibinfo{number}{4}
  (\bibinfo{year}{2007}), \bibinfo{pages}{462--473}.
\newblock


\bibitem[\protect\citeauthoryear{Card, Mackinlay, and Shneiderman}{Card
  et~al\mbox{.}}{1999}]%
        {card1999readings}
\bibfield{author}{\bibinfo{person}{Stuart~K Card}, \bibinfo{person}{Jock~D
  Mackinlay}, {and} \bibinfo{person}{Ben Shneiderman}.}
  \bibinfo{year}{1999}\natexlab{}.
\newblock \bibinfo{booktitle}{\emph{Readings in information visualization:
  using vision to think}}.
\newblock \bibinfo{publisher}{Morgan Kaufmann}.
\newblock


\bibitem[\protect\citeauthoryear{Carroll, Au, Detwiler, Fu, Painter, and
  Abernethy}{Carroll et~al\mbox{.}}{2014}]%
        {carroll2014visualization}
\bibfield{author}{\bibinfo{person}{Lauren~N Carroll}, \bibinfo{person}{Alan~P
  Au}, \bibinfo{person}{Landon~Todd Detwiler}, \bibinfo{person}{Tsung-chieh
  Fu}, \bibinfo{person}{Ian~S Painter}, {and} \bibinfo{person}{Neil~F
  Abernethy}.} \bibinfo{year}{2014}\natexlab{}.
\newblock \showarticletitle{Visualization and analytics tools for infectious
  disease epidemiology: a systematic review}.
\newblock \bibinfo{journal}{\emph{Journal of biomedical informatics}}
  \bibinfo{volume}{51} (\bibinfo{year}{2014}), \bibinfo{pages}{287--298}.
\newblock


\bibitem[\protect\citeauthoryear{Celi, Hassan, Marquardt, Breslow, and
  Rosenfeld}{Celi et~al\mbox{.}}{2001}]%
        {celi2001eicu}
\bibfield{author}{\bibinfo{person}{Leo~Anthony Celi}, \bibinfo{person}{Erkan
  Hassan}, \bibinfo{person}{Cynthia Marquardt}, \bibinfo{person}{Michael
  Breslow}, {and} \bibinfo{person}{Brian Rosenfeld}.}
  \bibinfo{year}{2001}\natexlab{}.
\newblock \showarticletitle{The eICU: it’s not just telemedicine}.
\newblock \bibinfo{journal}{\emph{Critical care medicine}}
  \bibinfo{volume}{29}, \bibinfo{number}{8} (\bibinfo{year}{2001}),
  \bibinfo{pages}{N183--N189}.
\newblock


\bibitem[\protect\citeauthoryear{Chen, Kumar, Fitch, Jagadish, Zhang, Dunn, and
  Chau}{Chen et~al\mbox{.}}{2015}]%
        {chen2015explicu}
\bibfield{author}{\bibinfo{person}{Robert Chen}, \bibinfo{person}{Vikas Kumar},
  \bibinfo{person}{Natalie Fitch}, \bibinfo{person}{Jitesh Jagadish},
  \bibinfo{person}{Lifan Zhang}, \bibinfo{person}{William Dunn}, {and}
  \bibinfo{person}{Duen~Horng Chau}.} \bibinfo{year}{2015}\natexlab{}.
\newblock \showarticletitle{explICU: A web-based visualization and predictive
  modeling toolkit for mortality in intensive care patients}. In
  \bibinfo{booktitle}{\emph{Engineering in Medicine and Biology Society (EMBC),
  2015 37th Annual International Conference of the IEEE}}. IEEE,
  \bibinfo{pages}{6830--6833}.
\newblock


\bibitem[\protect\citeauthoryear{Chou}{Chou}{2012}]%
        {chou2012health}
\bibfield{author}{\bibinfo{person}{David Chou}.}
  \bibinfo{year}{2012}\natexlab{}.
\newblock \showarticletitle{Health IT and patient safety: building safer
  systems for better care}.
\newblock \bibinfo{journal}{\emph{Jama}} \bibinfo{volume}{308},
  \bibinfo{number}{21} (\bibinfo{year}{2012}), \bibinfo{pages}{2282--2282}.
\newblock


\bibitem[\protect\citeauthoryear{Cohen, Blatter, Almeida, Shortliffe, and
  Patel}{Cohen et~al\mbox{.}}{2006}]%
        {cohen2006cognitive}
\bibfield{author}{\bibinfo{person}{Trevor Cohen}, \bibinfo{person}{Brett
  Blatter}, \bibinfo{person}{Carlos Almeida}, \bibinfo{person}{Edward
  Shortliffe}, {and} \bibinfo{person}{Vimla Patel}.}
  \bibinfo{year}{2006}\natexlab{}.
\newblock \showarticletitle{A cognitive blueprint of collaboration in context:
  Distributed cognition in the psychiatric emergency department}.
\newblock \bibinfo{journal}{\emph{Artificial intelligence in medicine}}
  \bibinfo{volume}{37}, \bibinfo{number}{2} (\bibinfo{year}{2006}),
  \bibinfo{pages}{73--83}.
\newblock


\bibitem[\protect\citeauthoryear{Colligan, Potts, Finn, and Sinkin}{Colligan
  et~al\mbox{.}}{2015}]%
        {colligan2015cognitive}
\bibfield{author}{\bibinfo{person}{Lacey Colligan}, \bibinfo{person}{Henry~WW
  Potts}, \bibinfo{person}{Chelsea~T Finn}, {and} \bibinfo{person}{Robert~A
  Sinkin}.} \bibinfo{year}{2015}\natexlab{}.
\newblock \showarticletitle{Cognitive workload changes for nurses transitioning
  from a legacy system with paper documentation to a commercial electronic
  health record}.
\newblock \bibinfo{journal}{\emph{International journal of medical
  informatics}} \bibinfo{volume}{84}, \bibinfo{number}{7}
  (\bibinfo{year}{2015}), \bibinfo{pages}{469--476}.
\newblock


\bibitem[\protect\citeauthoryear{Cook and Thomas}{Cook and Thomas}{2005}]%
        {cook2005illuminating}
\bibfield{author}{\bibinfo{person}{Kristin~A Cook} {and}
  \bibinfo{person}{James~J Thomas}.} \bibinfo{year}{2005}\natexlab{}.
\newblock \showarticletitle{Illuminating the path: The research and development
  agenda for visual analytics}.
\newblock  (\bibinfo{year}{2005}).
\newblock


\bibitem[\protect\citeauthoryear{Council et~al\mbox{.}}{Council
  et~al\mbox{.}}{2012}]%
        {national2012health}
\bibfield{author}{\bibinfo{person}{National~Research Council} {et~al\mbox{.}}}
  \bibinfo{year}{2012}\natexlab{}.
\newblock \bibinfo{title}{Health IT and patient safety: building safer systems
  for better care}.
\newblock
\newblock


\bibitem[\protect\citeauthoryear{Coustasse, Deslich, Bailey, Hairston, and
  Paul}{Coustasse et~al\mbox{.}}{2014}]%
        {coustasse2014business}
\bibfield{author}{\bibinfo{person}{Alberto Coustasse}, \bibinfo{person}{Stacie
  Deslich}, \bibinfo{person}{Deanna Bailey}, \bibinfo{person}{Alesia Hairston},
  {and} \bibinfo{person}{David Paul}.} \bibinfo{year}{2014}\natexlab{}.
\newblock \showarticletitle{A business case for tele-intensive care units}.
\newblock \bibinfo{journal}{\emph{The Permanente Journal}}
  \bibinfo{volume}{18}, \bibinfo{number}{4} (\bibinfo{year}{2014}),
  \bibinfo{pages}{76}.
\newblock


\bibitem[\protect\citeauthoryear{D'Aragon, Belley-Cote, Meade, Lauzier,
  Adhikari, Briel, Lalu, Kanji, Asfar, Turgeon, et~al\mbox{.}}{D'Aragon
  et~al\mbox{.}}{2015}]%
        {d2015blood}
\bibfield{author}{\bibinfo{person}{Frederick D'Aragon},
  \bibinfo{person}{Emilie~P Belley-Cote}, \bibinfo{person}{Maureen~O Meade},
  \bibinfo{person}{Fran{\c{c}}ois Lauzier}, \bibinfo{person}{Neill~KJ
  Adhikari}, \bibinfo{person}{Matthias Briel}, \bibinfo{person}{Manoj Lalu},
  \bibinfo{person}{Salmaan Kanji}, \bibinfo{person}{Pierre Asfar},
  \bibinfo{person}{Alexis~F Turgeon}, {et~al\mbox{.}}}
  \bibinfo{year}{2015}\natexlab{}.
\newblock \showarticletitle{Blood Pressure Targets For Vasopressor Therapy: A
  Systematic Review}.
\newblock \bibinfo{journal}{\emph{Shock}} \bibinfo{volume}{43},
  \bibinfo{number}{6} (\bibinfo{year}{2015}), \bibinfo{pages}{530--539}.
\newblock


\bibitem[\protect\citeauthoryear{Dix}{Dix}{2009}]%
        {dix2009human}
\bibfield{author}{\bibinfo{person}{Alan Dix}.} \bibinfo{year}{2009}\natexlab{}.
\newblock \showarticletitle{Human-computer interaction}.
\newblock In \bibinfo{booktitle}{\emph{Encyclopedia of database systems}}.
  \bibinfo{publisher}{Springer}, \bibinfo{pages}{1327--1331}.
\newblock


\bibitem[\protect\citeauthoryear{Ellsworth, Dziadzko, O’Horo, Farrell, Zhang,
  and Herasevich}{Ellsworth et~al\mbox{.}}{2016}]%
        {ellsworth2016appraisal}
\bibfield{author}{\bibinfo{person}{Marc~A Ellsworth}, \bibinfo{person}{Mikhail
  Dziadzko}, \bibinfo{person}{John~C O’Horo}, \bibinfo{person}{Ann~M
  Farrell}, \bibinfo{person}{Jiajie Zhang}, {and} \bibinfo{person}{Vitaly
  Herasevich}.} \bibinfo{year}{2016}\natexlab{}.
\newblock \showarticletitle{An appraisal of published usability evaluations of
  electronic health records via systematic review}.
\newblock \bibinfo{journal}{\emph{Journal of the American Medical Informatics
  Association}} \bibinfo{volume}{24}, \bibinfo{number}{1}
  (\bibinfo{year}{2016}), \bibinfo{pages}{218--226}.
\newblock


\bibitem[\protect\citeauthoryear{Elstein, Shulman, and Sprafka}{Elstein
  et~al\mbox{.}}{1978}]%
        {elstein1978medical}
\bibfield{author}{\bibinfo{person}{Arthur~S Elstein}, \bibinfo{person}{Lee~S
  Shulman}, {and} \bibinfo{person}{Sarah~A Sprafka}.}
  \bibinfo{year}{1978}\natexlab{}.
\newblock \showarticletitle{Medical problem solving an analysis of clinical
  reasoning}.
\newblock  (\bibinfo{year}{1978}).
\newblock


\bibitem[\protect\citeauthoryear{Faiola~A and Duke}{Faiola~A and Duke}{2015}]%
        {faiola2015}
\bibfield{author}{\bibinfo{person}{P Faiola~A, Srinivas} {and}
  \bibinfo{person}{J Duke}.} \bibinfo{year}{2015}\natexlab{}.
\newblock \showarticletitle{Supporting Clinical Cognition: A Human-Centered
  Approach to a Novel ICU Information Visualization Dashboard}. In
  \bibinfo{booktitle}{\emph{AMIA Annu Symp Proc}}. AMIA,
  \bibinfo{pages}{560--569}.
\newblock


\bibitem[\protect\citeauthoryear{Fischer}{Fischer}{2001}]%
        {fischer2001communities}
\bibfield{author}{\bibinfo{person}{Gerhard Fischer}.}
  \bibinfo{year}{2001}\natexlab{}.
\newblock \showarticletitle{Communities of interest: Learning through the
  interaction of multiple knowledge systems}. In
  \bibinfo{booktitle}{\emph{Proceedings of the 24th IRIS Conference}},
  Vol.~\bibinfo{volume}{1}. Department of Information Science, Bergen,
  \bibinfo{pages}{1--13}.
\newblock


\bibitem[\protect\citeauthoryear{Ghassemi, Naumann, Schulam, Beam, and
  Ranganath}{Ghassemi et~al\mbox{.}}{2018}]%
        {ghassemi2018opportunities}
\bibfield{author}{\bibinfo{person}{Marzyeh Ghassemi}, \bibinfo{person}{Tristan
  Naumann}, \bibinfo{person}{Peter Schulam}, \bibinfo{person}{Andrew~L Beam},
  {and} \bibinfo{person}{Rajesh Ranganath}.} \bibinfo{year}{2018}\natexlab{}.
\newblock \showarticletitle{Opportunities in Machine Learning for Healthcare}.
\newblock \bibinfo{journal}{\emph{arXiv preprint arXiv:1806.00388}}
  (\bibinfo{year}{2018}).
\newblock


\bibitem[\protect\citeauthoryear{Han, Carcillo, Venkataraman, Clark, Watson,
  Nguyen, Bayir, and Orr}{Han et~al\mbox{.}}{2005}]%
        {han2005unexpected}
\bibfield{author}{\bibinfo{person}{Yong~Y Han}, \bibinfo{person}{Joseph~A
  Carcillo}, \bibinfo{person}{Shekhar~T Venkataraman},
  \bibinfo{person}{Robert~SB Clark}, \bibinfo{person}{R~Scott Watson},
  \bibinfo{person}{Trung~C Nguyen}, \bibinfo{person}{H{\"u}lya Bayir}, {and}
  \bibinfo{person}{Richard~A Orr}.} \bibinfo{year}{2005}\natexlab{}.
\newblock \showarticletitle{Unexpected increased mortality after implementation
  of a commercially sold computerized physician order entry system}.
\newblock \bibinfo{journal}{\emph{Pediatrics}} \bibinfo{volume}{116},
  \bibinfo{number}{6} (\bibinfo{year}{2005}), \bibinfo{pages}{1506--1512}.
\newblock


\bibitem[\protect\citeauthoryear{Hart and Staveland}{Hart and
  Staveland}{1988}]%
        {hart1988development}
\bibfield{author}{\bibinfo{person}{Sandra~G Hart} {and}
  \bibinfo{person}{Lowell~E Staveland}.} \bibinfo{year}{1988}\natexlab{}.
\newblock \showarticletitle{Development of NASA-TLX (Task Load Index): Results
  of empirical and theoretical research}.
\newblock In \bibinfo{booktitle}{\emph{Advances in psychology}}.
  Vol.~\bibinfo{volume}{52}. \bibinfo{publisher}{Elsevier},
  \bibinfo{pages}{139--183}.
\newblock


\bibitem[\protect\citeauthoryear{H{\"a}yrinen, Saranto, and
  Nyk{\"a}nen}{H{\"a}yrinen et~al\mbox{.}}{2008}]%
        {Hyrinen2008DefinitionSC}
\bibfield{author}{\bibinfo{person}{Kristiina H{\"a}yrinen},
  \bibinfo{person}{Kaija Saranto}, {and} \bibinfo{person}{Pirkko Nyk{\"a}nen}.}
  \bibinfo{year}{2008}\natexlab{}.
\newblock \showarticletitle{Definition, structure, content, use and impacts of
  electronic health records: A review of the research literature}.
\newblock \bibinfo{journal}{\emph{International journal of medical
  informatics}}  \bibinfo{volume}{77 5} (\bibinfo{year}{2008}),
  \bibinfo{pages}{291--304}.
\newblock


\bibitem[\protect\citeauthoryear{Heath and Luff}{Heath and Luff}{1996}]%
        {heath1996documents}
\bibfield{author}{\bibinfo{person}{Christian Heath} {and} \bibinfo{person}{Paul
  Luff}.} \bibinfo{year}{1996}\natexlab{}.
\newblock \showarticletitle{Documents and professional practice:“bad”
  organisational reasons for “good” clinical records}. In
  \bibinfo{booktitle}{\emph{Proceedings of the 1996 ACM conference on Computer
  supported cooperative work}}. ACM, \bibinfo{pages}{354--363}.
\newblock


\bibitem[\protect\citeauthoryear{Henriksen, Dayton, Keyes, Carayon, and
  Hughes}{Henriksen et~al\mbox{.}}{2008}]%
        {henriksen2008understanding}
\bibfield{author}{\bibinfo{person}{Kerm Henriksen}, \bibinfo{person}{Elizabeth
  Dayton}, \bibinfo{person}{Margaret~A Keyes}, \bibinfo{person}{Pascale
  Carayon}, {and} \bibinfo{person}{Ronda Hughes}.}
  \bibinfo{year}{2008}\natexlab{}.
\newblock \showarticletitle{Understanding adverse events: a human factors
  framework}.
\newblock  (\bibinfo{year}{2008}).
\newblock


\bibitem[\protect\citeauthoryear{Holzinger}{Holzinger}{2005}]%
        {holzinger2005usability}
\bibfield{author}{\bibinfo{person}{Andreas Holzinger}.}
  \bibinfo{year}{2005}\natexlab{}.
\newblock \showarticletitle{Usability engineering methods for software
  developers}.
\newblock \bibinfo{journal}{\emph{Commun. ACM}} \bibinfo{volume}{48},
  \bibinfo{number}{1} (\bibinfo{year}{2005}), \bibinfo{pages}{71--74}.
\newblock


\bibitem[\protect\citeauthoryear{Horowitz, Altaie, and Boyd}{Horowitz
  et~al\mbox{.}}{2010}]%
        {horowitz2010defining}
\bibfield{author}{\bibinfo{person}{Gary~L Horowitz}, \bibinfo{person}{Sousan
  Altaie}, {and} \bibinfo{person}{James~C Boyd}.}
  \bibinfo{year}{2010}\natexlab{}.
\newblock \bibinfo{booktitle}{\emph{Defining, establishing, and verifying
  reference intervals in the clinical laboratory; approved guideline}}.
\newblock \bibinfo{publisher}{CLSI}.
\newblock


\bibitem[\protect\citeauthoryear{JL, KT, A, and RM}{JL et~al\mbox{.}}{2018}]%
        {doi:10.1001/jama.2018.1171}
\bibfield{author}{\bibinfo{person}{Howe JL}, \bibinfo{person}{Adams KT},
  \bibinfo{person}{Hettinger A}, {and} \bibinfo{person}{Ratwani RM}.}
  \bibinfo{year}{2018}\natexlab{}.
\newblock \showarticletitle{Electronic health record usability issues and
  potential contribution to patient harm}.
\newblock \bibinfo{journal}{\emph{JAMA}} \bibinfo{volume}{319},
  \bibinfo{number}{12} (\bibinfo{year}{2018}), \bibinfo{pages}{1276--1278}.
\newblock
\urldef\tempurl%
\url{https://doi.org/10.1001/jama.2018.1171}
\showDOI{\tempurl}


\bibitem[\protect\citeauthoryear{Johnson, Pollard, Shen, Lehman, Feng,
  Ghassemi, Moody, Szolovits, Celi, and Mark}{Johnson et~al\mbox{.}}{2016}]%
        {johnson2016mimiciii}
\bibfield{author}{\bibinfo{person}{Alistair~EW Johnson}, \bibinfo{person}{Tom~J
  Pollard}, \bibinfo{person}{Lu Shen}, \bibinfo{person}{Li-wei~H Lehman},
  \bibinfo{person}{Mengling Feng}, \bibinfo{person}{Mohammad Ghassemi},
  \bibinfo{person}{Benjamin Moody}, \bibinfo{person}{Peter Szolovits},
  \bibinfo{person}{Leo~Anthony Celi}, {and} \bibinfo{person}{Roger~G Mark}.}
  \bibinfo{year}{2016}\natexlab{}.
\newblock \showarticletitle{{MIMIC-III}, a freely accessible critical care
  database}.
\newblock \bibinfo{journal}{\emph{Scientific data}}  \bibinfo{volume}{3}
  (\bibinfo{year}{2016}).
\newblock


\bibitem[\protect\citeauthoryear{Keedwell and D{\'e}nes}{Keedwell and
  D{\'e}nes}{2015}]%
        {keedwell2015latin}
\bibfield{author}{\bibinfo{person}{A~Donald Keedwell} {and}
  \bibinfo{person}{J{\'o}zsef D{\'e}nes}.} \bibinfo{year}{2015}\natexlab{}.
\newblock \bibinfo{booktitle}{\emph{Latin squares and their applications}}.
\newblock \bibinfo{publisher}{Elsevier}.
\newblock


\bibitem[\protect\citeauthoryear{Kenyon}{Kenyon}{[n. d.]}]%
        {Kenyon1}
\bibfield{author}{\bibinfo{person}{Kathy Kenyon}.} \bibinfo{year}{[n.
  d.]}\natexlab{}.
\newblock \bibinfo{title}{Overcoming Contractual Barriers To EHR Research}.
\newblock
\newblock
\urldef\tempurl%
\url{https://www.healthaffairs.org/do/10.1377/hblog20151014.051141/full/}
\showURL{%
\tempurl}


\bibitem[\protect\citeauthoryear{Khairat, Coleman, Russomagno, and
  Gotz}{Khairat et~al\mbox{.}}{[n. d.]}]%
        {khairatassessing}
\bibfield{author}{\bibinfo{person}{Saif Khairat},
  \bibinfo{person}{George~Cameron Coleman}, \bibinfo{person}{Samantha
  Russomagno}, {and} \bibinfo{person}{David Gotz}.} \bibinfo{year}{[n.
  d.]}\natexlab{}.
\newblock \showarticletitle{Assessing the Status Quo of EHR Accessibility,
  Usability, and Knowledge Dissemination}.
\newblock  (\bibinfo{year}{[n. d.]}).
\newblock


\bibitem[\protect\citeauthoryear{Klasnja, Civan~Hartzler, Unruh, and
  Pratt}{Klasnja et~al\mbox{.}}{2010}]%
        {klasnja2010blowing}
\bibfield{author}{\bibinfo{person}{Predrag Klasnja}, \bibinfo{person}{Andrea
  Civan~Hartzler}, \bibinfo{person}{Kent~T Unruh}, {and} \bibinfo{person}{Wanda
  Pratt}.} \bibinfo{year}{2010}\natexlab{}.
\newblock \showarticletitle{Blowing in the wind: unanchored patient information
  work during cancer care}. In \bibinfo{booktitle}{\emph{Proceedings of the
  SIGCHI Conference on Human Factors in Computing Systems}}. ACM,
  \bibinfo{pages}{193--202}.
\newblock


\bibitem[\protect\citeauthoryear{Klimov, Shahar, and Taieb-Maimon}{Klimov
  et~al\mbox{.}}{2010}]%
        {klimov2010intelligent}
\bibfield{author}{\bibinfo{person}{Denis Klimov}, \bibinfo{person}{Yuval
  Shahar}, {and} \bibinfo{person}{Meirav Taieb-Maimon}.}
  \bibinfo{year}{2010}\natexlab{}.
\newblock \showarticletitle{Intelligent visualization and exploration of
  time-oriented data of multiple patients}.
\newblock \bibinfo{journal}{\emph{Artificial intelligence in medicine}}
  \bibinfo{volume}{49}, \bibinfo{number}{1} (\bibinfo{year}{2010}),
  \bibinfo{pages}{11--31}.
\newblock


\bibitem[\protect\citeauthoryear{Lee, Ribey, and Wallace}{Lee
  et~al\mbox{.}}{2015}]%
        {lee2015web}
\bibfield{author}{\bibinfo{person}{Joon Lee}, \bibinfo{person}{Evan Ribey},
  {and} \bibinfo{person}{James~R Wallace}.} \bibinfo{year}{2015}\natexlab{}.
\newblock \showarticletitle{A web-based data visualization tool for the
  MIMIC-II database}.
\newblock \bibinfo{journal}{\emph{BMC medical informatics and decision making}}
  \bibinfo{volume}{16}, \bibinfo{number}{1} (\bibinfo{year}{2015}),
  \bibinfo{pages}{15}.
\newblock


\bibitem[\protect\citeauthoryear{Levy-Lambert, Gong, Naumann, Pollard, and
  Guttag}{Levy-Lambert et~al\mbox{.}}{2018}]%
        {levy2018visualizing}
\bibfield{author}{\bibinfo{person}{Dina Levy-Lambert}, \bibinfo{person}{Jen~J
  Gong}, \bibinfo{person}{Tristan Naumann}, \bibinfo{person}{Tom~J Pollard},
  {and} \bibinfo{person}{John~V Guttag}.} \bibinfo{year}{2018}\natexlab{}.
\newblock \showarticletitle{Visualizing Patient Timelines in the Intensive Care
  Unit}.
\newblock \bibinfo{journal}{\emph{arXiv preprint arXiv:1806.00397}}
  (\bibinfo{year}{2018}).
\newblock


\bibitem[\protect\citeauthoryear{Lewis and Rieman}{Lewis and Rieman}{1993}]%
        {lewis1993task}
\bibfield{author}{\bibinfo{person}{Clayton Lewis} {and} \bibinfo{person}{John
  Rieman}.} \bibinfo{year}{1993}\natexlab{}.
\newblock \showarticletitle{Task-centered user interface design}.
\newblock \bibinfo{journal}{\emph{A Practical Introduction}}
  (\bibinfo{year}{1993}).
\newblock


\bibitem[\protect\citeauthoryear{Massey~Jr}{Massey~Jr}{1951}]%
        {massey1951kolmogorov}
\bibfield{author}{\bibinfo{person}{Frank~J Massey~Jr}.}
  \bibinfo{year}{1951}\natexlab{}.
\newblock \showarticletitle{The Kolmogorov-Smirnov test for goodness of fit}.
\newblock \bibinfo{journal}{\emph{Journal of the American statistical
  Association}} \bibinfo{volume}{46}, \bibinfo{number}{253}
  (\bibinfo{year}{1951}), \bibinfo{pages}{68--78}.
\newblock


\bibitem[\protect\citeauthoryear{Mazur, Mosaly, Moore, Comitz, Yu, Falchook,
  Eblan, Hoyle, Tracton, Chera, et~al\mbox{.}}{Mazur et~al\mbox{.}}{2016}]%
        {mazur2016toward}
\bibfield{author}{\bibinfo{person}{Lukasz~M Mazur}, \bibinfo{person}{Prithima~R
  Mosaly}, \bibinfo{person}{Carlton Moore}, \bibinfo{person}{Elizabeth Comitz},
  \bibinfo{person}{Fei Yu}, \bibinfo{person}{Aaron~D Falchook},
  \bibinfo{person}{Michael~J Eblan}, \bibinfo{person}{Lesley~M Hoyle},
  \bibinfo{person}{Gregg Tracton}, \bibinfo{person}{Bhishamjit~S Chera},
  {et~al\mbox{.}}} \bibinfo{year}{2016}\natexlab{}.
\newblock \showarticletitle{Toward a better understanding of task demands,
  workload, and performance during physician-computer interactions}.
\newblock \bibinfo{journal}{\emph{Journal of the American Medical Informatics
  Association}} \bibinfo{volume}{23}, \bibinfo{number}{6}
  (\bibinfo{year}{2016}), \bibinfo{pages}{1113--1120}.
\newblock


\bibitem[\protect\citeauthoryear{Middleton, Sittig, and Wright}{Middleton
  et~al\mbox{.}}{2016}]%
        {middleton2016clinical}
\bibfield{author}{\bibinfo{person}{B Middleton}, \bibinfo{person}{DF Sittig},
  {and} \bibinfo{person}{A Wright}.} \bibinfo{year}{2016}\natexlab{}.
\newblock \showarticletitle{Clinical Decision Support: a 25 Year Retrospective
  and a 25 Year Vision.}
\newblock \bibinfo{journal}{\emph{Yearbook of medical informatics}}
  (\bibinfo{year}{2016}), \bibinfo{pages}{S103}.
\newblock


\bibitem[\protect\citeauthoryear{Mosaly, Mazur, Yu, Guo, Derek, Laidlaw, Moore,
  Marks, and Mostafa}{Mosaly et~al\mbox{.}}{2018}]%
        {mosaly2018relating}
\bibfield{author}{\bibinfo{person}{Prithima~Reddy Mosaly},
  \bibinfo{person}{Lukasz~M Mazur}, \bibinfo{person}{Fei Yu},
  \bibinfo{person}{Hua Guo}, \bibinfo{person}{Merck Derek},
  \bibinfo{person}{David~H Laidlaw}, \bibinfo{person}{Carlton Moore},
  \bibinfo{person}{Lawrence~B Marks}, {and} \bibinfo{person}{Javed Mostafa}.}
  \bibinfo{year}{2018}\natexlab{}.
\newblock \showarticletitle{Relating task demand, mental effort and task
  difficulty with physicians' performance during interactions with electronic
  health records (EHRs)}.
\newblock \bibinfo{journal}{\emph{International Journal of Human--Computer
  Interaction}} \bibinfo{volume}{34}, \bibinfo{number}{5}
  (\bibinfo{year}{2018}), \bibinfo{pages}{467--475}.
\newblock


\bibitem[\protect\citeauthoryear{M{\"u}llner, Urbanek, Havel, Losert, Gamper,
  and Herkner}{M{\"u}llner et~al\mbox{.}}{2004}]%
        {mullner2004vasopressors}
\bibfield{author}{\bibinfo{person}{Marcus M{\"u}llner},
  \bibinfo{person}{Bernhard Urbanek}, \bibinfo{person}{Christof Havel},
  \bibinfo{person}{Heidrun Losert}, \bibinfo{person}{Gunnar Gamper}, {and}
  \bibinfo{person}{Harald Herkner}.} \bibinfo{year}{2004}\natexlab{}.
\newblock \showarticletitle{Vasopressors for shock}.
\newblock \bibinfo{journal}{\emph{The Cochrane Library}}
  (\bibinfo{year}{2004}).
\newblock


\bibitem[\protect\citeauthoryear{Nolan, Cartin-Ceba, Moreno-Franco, Pickering,
  and Herasevich}{Nolan et~al\mbox{.}}{2017a}]%
        {nolan2017multisite}
\bibfield{author}{\bibinfo{person}{Matthew~E Nolan}, \bibinfo{person}{Rodrigo
  Cartin-Ceba}, \bibinfo{person}{Pablo Moreno-Franco}, \bibinfo{person}{Brian
  Pickering}, {and} \bibinfo{person}{Vitaly Herasevich}.}
  \bibinfo{year}{2017}\natexlab{a}.
\newblock \showarticletitle{A Multisite Survey Study of EMR Review Habits,
  Information Needs, and Display Preferences among Medical ICU Clinicians
  Evaluating New Patients}.
\newblock \bibinfo{journal}{\emph{Applied clinical informatics}}
  \bibinfo{volume}{8}, \bibinfo{number}{04} (\bibinfo{year}{2017}),
  \bibinfo{pages}{1197--1207}.
\newblock


\bibitem[\protect\citeauthoryear{Nolan, Siwani, Helmi, Pickering,
  Moreno-Franco, and Herasevich}{Nolan et~al\mbox{.}}{2017b}]%
        {nolan2017health}
\bibfield{author}{\bibinfo{person}{Matthew~E Nolan}, \bibinfo{person}{Rizwan
  Siwani}, \bibinfo{person}{Haytham Helmi}, \bibinfo{person}{Brian~W
  Pickering}, \bibinfo{person}{Pablo Moreno-Franco}, {and}
  \bibinfo{person}{Vitaly Herasevich}.} \bibinfo{year}{2017}\natexlab{b}.
\newblock \showarticletitle{Health IT Usability Focus Section: Data Use and
  Navigation Patterns among Medical ICU Clinicians during Electronic Chart
  Review}.
\newblock \bibinfo{journal}{\emph{Applied clinical informatics}}
  \bibinfo{volume}{8}, \bibinfo{number}{04} (\bibinfo{year}{2017}),
  \bibinfo{pages}{1117--1126}.
\newblock


\bibitem[\protect\citeauthoryear{Norman}{Norman}{2009}]%
        {norman2009dual}
\bibfield{author}{\bibinfo{person}{Geoff Norman}.}
  \bibinfo{year}{2009}\natexlab{}.
\newblock \showarticletitle{Dual processing and diagnostic errors}.
\newblock \bibinfo{journal}{\emph{Advances in Health Sciences Education}}
  \bibinfo{volume}{14}, \bibinfo{number}{1} (\bibinfo{year}{2009}),
  \bibinfo{pages}{37--49}.
\newblock


\bibitem[\protect\citeauthoryear{Olchanski, Dziadzko, Tiong, Daniels, Peters,
  O'Horo, and Gong}{Olchanski et~al\mbox{.}}{2017}]%
        {olchanski2017can}
\bibfield{author}{\bibinfo{person}{Natalia Olchanski},
  \bibinfo{person}{Mikhail~A Dziadzko}, \bibinfo{person}{C Tiong},
  \bibinfo{person}{Craig~E Daniels}, \bibinfo{person}{Steve~G Peters},
  \bibinfo{person}{John~C O'Horo}, {and} \bibinfo{person}{Michelle~N Gong}.}
  \bibinfo{year}{2017}\natexlab{}.
\newblock \showarticletitle{Can a Novel ICU Data Display Positively Affect
  Patient Outcomes and Save Lives?}
\newblock \bibinfo{journal}{\emph{Journal of medical systems}}
  \bibinfo{volume}{41}, \bibinfo{number}{11} (\bibinfo{year}{2017}),
  \bibinfo{pages}{171}.
\newblock


\bibitem[\protect\citeauthoryear{O’malley, Grossman, Cohen, Kemper, and
  Pham}{O’malley et~al\mbox{.}}{2010}]%
        {o2010electronic}
\bibfield{author}{\bibinfo{person}{Ann~S O’malley}, \bibinfo{person}{Joy~M
  Grossman}, \bibinfo{person}{Genna~R Cohen}, \bibinfo{person}{Nicole~M
  Kemper}, {and} \bibinfo{person}{Hoangmai~H Pham}.}
  \bibinfo{year}{2010}\natexlab{}.
\newblock \showarticletitle{Are electronic medical records helpful for care
  coordination? Experiences of physician practices}.
\newblock \bibinfo{journal}{\emph{Journal of general internal medicine}}
  \bibinfo{volume}{25}, \bibinfo{number}{3} (\bibinfo{year}{2010}),
  \bibinfo{pages}{177--185}.
\newblock


\bibitem[\protect\citeauthoryear{Pickering, Dong, Ahmed, Giri, Kilickaya,
  Gupta, Gajic, and Herasevich}{Pickering et~al\mbox{.}}{2015}]%
        {pickering2015implementation}
\bibfield{author}{\bibinfo{person}{Brian~W Pickering}, \bibinfo{person}{Yue
  Dong}, \bibinfo{person}{Adil Ahmed}, \bibinfo{person}{Jyothsna Giri},
  \bibinfo{person}{Oguz Kilickaya}, \bibinfo{person}{Ashish Gupta},
  \bibinfo{person}{Ognjen Gajic}, {and} \bibinfo{person}{Vitaly Herasevich}.}
  \bibinfo{year}{2015}\natexlab{}.
\newblock \showarticletitle{The implementation of clinician designed,
  human-centered electronic medical record viewer in the intensive care unit: a
  pilot step-wedge cluster randomized trial}.
\newblock \bibinfo{journal}{\emph{International journal of medical
  informatics}} \bibinfo{volume}{84}, \bibinfo{number}{5}
  (\bibinfo{year}{2015}), \bibinfo{pages}{299--307}.
\newblock


\bibitem[\protect\citeauthoryear{Plaisant, Milash, Rose, Widoff, and
  Shneiderman}{Plaisant et~al\mbox{.}}{1996}]%
        {plaisant1996lifelines}
\bibfield{author}{\bibinfo{person}{Catherine Plaisant}, \bibinfo{person}{Brett
  Milash}, \bibinfo{person}{Anne Rose}, \bibinfo{person}{Seth Widoff}, {and}
  \bibinfo{person}{Ben Shneiderman}.} \bibinfo{year}{1996}\natexlab{}.
\newblock \showarticletitle{LifeLines: visualizing personal histories}. In
  \bibinfo{booktitle}{\emph{Proceedings of the SIGCHI conference on Human
  factors in computing systems}}. ACM, \bibinfo{pages}{221--227}.
\newblock


\bibitem[\protect\citeauthoryear{Rajkomar and Dhaliwal}{Rajkomar and
  Dhaliwal}{2011}]%
        {rajkomar2011improving}
\bibfield{author}{\bibinfo{person}{Alvin Rajkomar} {and}
  \bibinfo{person}{Gurpreet Dhaliwal}.} \bibinfo{year}{2011}\natexlab{}.
\newblock \showarticletitle{Improving diagnostic reasoning to improve patient
  safety}.
\newblock \bibinfo{journal}{\emph{The Permanente Journal}}
  \bibinfo{volume}{15}, \bibinfo{number}{3} (\bibinfo{year}{2011}),
  \bibinfo{pages}{68}.
\newblock


\bibitem[\protect\citeauthoryear{Roman, Ancker, Johnson, and
  Senathirajah}{Roman et~al\mbox{.}}{2017}]%
        {roman2017navigation}
\bibfield{author}{\bibinfo{person}{Lisette~C Roman}, \bibinfo{person}{Jessica~S
  Ancker}, \bibinfo{person}{Stephen~B Johnson}, {and} \bibinfo{person}{Yalini
  Senathirajah}.} \bibinfo{year}{2017}\natexlab{}.
\newblock \showarticletitle{Navigation in the electronic health record: a
  review of the safety and usability literature}.
\newblock \bibinfo{journal}{\emph{Journal of biomedical informatics}}
  \bibinfo{volume}{67} (\bibinfo{year}{2017}), \bibinfo{pages}{69--79}.
\newblock


\bibitem[\protect\citeauthoryear{Rothschild, Landrigan, Cronin, Kaushal,
  Lockley, Burdick, Stone, Lilly, Katz, Czeisler, et~al\mbox{.}}{Rothschild
  et~al\mbox{.}}{2005}]%
        {rothschild2005critical}
\bibfield{author}{\bibinfo{person}{Jeffrey~M Rothschild},
  \bibinfo{person}{Christopher~P Landrigan}, \bibinfo{person}{John~W Cronin},
  \bibinfo{person}{Rainu Kaushal}, \bibinfo{person}{Steven~W Lockley},
  \bibinfo{person}{Elisabeth Burdick}, \bibinfo{person}{Peter~H Stone},
  \bibinfo{person}{Craig~M Lilly}, \bibinfo{person}{Joel~T Katz},
  \bibinfo{person}{Charles~A Czeisler}, {et~al\mbox{.}}}
  \bibinfo{year}{2005}\natexlab{}.
\newblock \showarticletitle{The Critical Care Safety Study: The incidence and
  nature of adverse events and serious medical errors in intensive care}.
\newblock \bibinfo{journal}{\emph{Critical care medicine}}
  \bibinfo{volume}{33}, \bibinfo{number}{8} (\bibinfo{year}{2005}),
  \bibinfo{pages}{1694--1700}.
\newblock


\bibitem[\protect\citeauthoryear{Savikhin, Maciejewski, and Ebert}{Savikhin
  et~al\mbox{.}}{2008}]%
        {savikhin2008applied}
\bibfield{author}{\bibinfo{person}{Anya Savikhin}, \bibinfo{person}{Ross
  Maciejewski}, {and} \bibinfo{person}{David~S Ebert}.}
  \bibinfo{year}{2008}\natexlab{}.
\newblock \showarticletitle{Applied visual analytics for economic
  decision-making}. In \bibinfo{booktitle}{\emph{Visual Analytics Science and
  Technology, 2008. Vast'08. Ieee Symposium on}}. IEEE,
  \bibinfo{pages}{107--114}.
\newblock


\bibitem[\protect\citeauthoryear{Sedlmair, Meyer, and Munzner}{Sedlmair
  et~al\mbox{.}}{2012}]%
        {sedlmair2012design}
\bibfield{author}{\bibinfo{person}{Michael Sedlmair}, \bibinfo{person}{Miriah
  Meyer}, {and} \bibinfo{person}{Tamara Munzner}.}
  \bibinfo{year}{2012}\natexlab{}.
\newblock \showarticletitle{Design study methodology: Reflections from the
  trenches and the stacks}.
\newblock \bibinfo{journal}{\emph{IEEE transactions on visualization and
  computer graphics}} \bibinfo{volume}{18}, \bibinfo{number}{12}
  (\bibinfo{year}{2012}), \bibinfo{pages}{2431--2440}.
\newblock


\bibitem[\protect\citeauthoryear{Shen-Hsieh and Schindl}{Shen-Hsieh and
  Schindl}{2002}]%
        {shen2002data}
\bibfield{author}{\bibinfo{person}{Angela Shen-Hsieh} {and}
  \bibinfo{person}{Mark Schindl}.} \bibinfo{year}{2002}\natexlab{}.
\newblock \showarticletitle{Data visualization for strategic decision making}.
  In \bibinfo{booktitle}{\emph{Case Studies of the CHI2002}}. ACM,
  \bibinfo{pages}{1--17}.
\newblock


\bibitem[\protect\citeauthoryear{Shneiderman, Plaisant, and Hesse}{Shneiderman
  et~al\mbox{.}}{2013}]%
        {shneiderman2013improving}
\bibfield{author}{\bibinfo{person}{Ben Shneiderman}, \bibinfo{person}{Catherine
  Plaisant}, {and} \bibinfo{person}{Bradford~W Hesse}.}
  \bibinfo{year}{2013}\natexlab{}.
\newblock \showarticletitle{Improving healthcare with interactive
  visualization}.
\newblock \bibinfo{journal}{\emph{Computer}} \bibinfo{volume}{46},
  \bibinfo{number}{5} (\bibinfo{year}{2013}), \bibinfo{pages}{58--66}.
\newblock


\bibitem[\protect\citeauthoryear{Sinsky, Hess, Karsh, Keller, and
  Koppel}{Sinsky et~al\mbox{.}}{2012}]%
        {sinsky2012comparative}
\bibfield{author}{\bibinfo{person}{C Sinsky}, \bibinfo{person}{J Hess},
  \bibinfo{person}{BT Karsh}, \bibinfo{person}{JP Keller}, {and}
  \bibinfo{person}{R Koppel}.} \bibinfo{year}{2012}\natexlab{}.
\newblock \showarticletitle{Comparative user experiences of health IT products:
  how user experiences would be reported and used}.
\newblock \bibinfo{journal}{\emph{Institute of Medicine of the National
  Academies}} (\bibinfo{year}{2012}).
\newblock


\bibitem[\protect\citeauthoryear{Steele, Cakiroglu, Shah, Denaxas, Hemingway,
  and Luscombe}{Steele et~al\mbox{.}}{2018}]%
        {steele2018machine}
\bibfield{author}{\bibinfo{person}{Andrew~J Steele}, \bibinfo{person}{S~Aylin
  Cakiroglu}, \bibinfo{person}{Anoop~D Shah}, \bibinfo{person}{Spiros~C
  Denaxas}, \bibinfo{person}{Harry Hemingway}, {and}
  \bibinfo{person}{Nicholas~M Luscombe}.} \bibinfo{year}{2018}\natexlab{}.
\newblock \showarticletitle{Machine learning models in electronic health
  records can outperform conventional survival models for predicting patient
  mortality in coronary artery disease}.
\newblock \bibinfo{journal}{\emph{bioRxiv}} (\bibinfo{year}{2018}),
  \bibinfo{pages}{256008}.
\newblock


\bibitem[\protect\citeauthoryear{Sutcliffe, Lewton, and Rosenthal}{Sutcliffe
  et~al\mbox{.}}{2004}]%
        {sutcliffe2004communication}
\bibfield{author}{\bibinfo{person}{Kathleen~M Sutcliffe},
  \bibinfo{person}{Elizabeth Lewton}, {and} \bibinfo{person}{Marilynn~M
  Rosenthal}.} \bibinfo{year}{2004}\natexlab{}.
\newblock \showarticletitle{Communication failures: an insidious contributor to
  medical mishaps}.
\newblock \bibinfo{journal}{\emph{Academic Medicine}} \bibinfo{volume}{79},
  \bibinfo{number}{2} (\bibinfo{year}{2004}), \bibinfo{pages}{186--194}.
\newblock


\bibitem[\protect\citeauthoryear{Tahir}{Tahir}{[n. d.]}]%
        {Tahir1}
\bibfield{author}{\bibinfo{person}{Darius Tahir}.} \bibinfo{year}{[n.
  d.]}\natexlab{}.
\newblock \bibinfo{title}{Doctors barred from discussing safety glitches in
  U.S.-funded software}.
\newblock
\newblock
\urldef\tempurl%
\url{https://www.politico.com/story/2015/09/doctors-barred-from-discussing-safety-glitches-in-us-funded-software-213553}
\showURL{%
\tempurl}


\bibitem[\protect\citeauthoryear{Tobin}{Tobin}{2006}]%
        {tobin2006principles}
\bibfield{editor}{\bibinfo{person}{Martin~J Tobin}} (Ed.).
  \bibinfo{year}{2006}\natexlab{}.
\newblock \bibinfo{booktitle}{\emph{Principles and practice of mechanical
  ventilation}}.
\newblock \bibinfo{publisher}{McGraw-Hill Medical Pub. Division}.
\newblock


\bibitem[\protect\citeauthoryear{Weber, Mandl, and Kohane}{Weber
  et~al\mbox{.}}{2014}]%
        {weber2014finding}
\bibfield{author}{\bibinfo{person}{Griffin~M Weber}, \bibinfo{person}{Kenneth~D
  Mandl}, {and} \bibinfo{person}{Isaac~S Kohane}.}
  \bibinfo{year}{2014}\natexlab{}.
\newblock \showarticletitle{Finding the missing link for big biomedical data}.
\newblock \bibinfo{journal}{\emph{Jama}} \bibinfo{volume}{311},
  \bibinfo{number}{24} (\bibinfo{year}{2014}), \bibinfo{pages}{2479--2480}.
\newblock


\bibitem[\protect\citeauthoryear{Wu, Roy, and Stewart}{Wu
  et~al\mbox{.}}{2010}]%
        {wu2010prediction}
\bibfield{author}{\bibinfo{person}{Jionglin Wu}, \bibinfo{person}{Jason Roy},
  {and} \bibinfo{person}{Walter~F Stewart}.} \bibinfo{year}{2010}\natexlab{}.
\newblock \showarticletitle{Prediction modeling using EHR data: challenges,
  strategies, and a comparison of machine learning approaches}.
\newblock \bibinfo{journal}{\emph{Medical care}} (\bibinfo{year}{2010}),
  \bibinfo{pages}{S106--S113}.
\newblock


\bibitem[\protect\citeauthoryear{Yang and Tobin}{Yang and Tobin}{1991}]%
        {yang1991prospective}
\bibfield{author}{\bibinfo{person}{Karl~L Yang} {and} \bibinfo{person}{Martin~J
  Tobin}.} \bibinfo{year}{1991}\natexlab{}.
\newblock \showarticletitle{A prospective study of indexes predicting the
  outcome of trials of weaning from mechanical ventilation}.
\newblock \bibinfo{journal}{\emph{New England Journal of Medicine}}
  \bibinfo{volume}{324}, \bibinfo{number}{21} (\bibinfo{year}{1991}),
  \bibinfo{pages}{1445--1450}.
\newblock


\bibitem[\protect\citeauthoryear{Zhang, Patel, Johnson, and Smith}{Zhang
  et~al\mbox{.}}{2002}]%
        {zhang2002designing}
\bibfield{author}{\bibinfo{person}{Jiajie Zhang}, \bibinfo{person}{Vimla~L
  Patel}, \bibinfo{person}{Kathy~A Johnson}, {and} \bibinfo{person}{Jack~W
  Smith}.} \bibinfo{year}{2002}\natexlab{}.
\newblock \showarticletitle{Designing human-centered distributed information
  systems}.
\newblock \bibinfo{journal}{\emph{IEEE intelligent systems}}
  \bibinfo{volume}{17}, \bibinfo{number}{5} (\bibinfo{year}{2002}),
  \bibinfo{pages}{42--47}.
\newblock


\bibitem[\protect\citeauthoryear{Zhou, Ackerman, and Zheng}{Zhou
  et~al\mbox{.}}{2009}]%
        {zhou2009just}
\bibfield{author}{\bibinfo{person}{Xiaomu Zhou}, \bibinfo{person}{Mark~S
  Ackerman}, {and} \bibinfo{person}{Kai Zheng}.}
  \bibinfo{year}{2009}\natexlab{}.
\newblock \showarticletitle{I just don't know why it's gone: maintaining
  informal information use in inpatient care}. In
  \bibinfo{booktitle}{\emph{Proceedings of the SIGCHI Conference on Human
  Factors in Computing Systems}}. ACM, \bibinfo{pages}{2061--2070}.
\newblock


\end{thebibliography}

\clearpage

\appendix
%Appendix A

\section{Baseline Overview}
\label{sec:baselineGUIAppendix}
The baseline visualization is a composite representation of a range of real commercial EHRs that the research team have directly observed.
\begin{figure*}[!ht]
  \centering
  \includegraphics[width=0.75\linewidth]{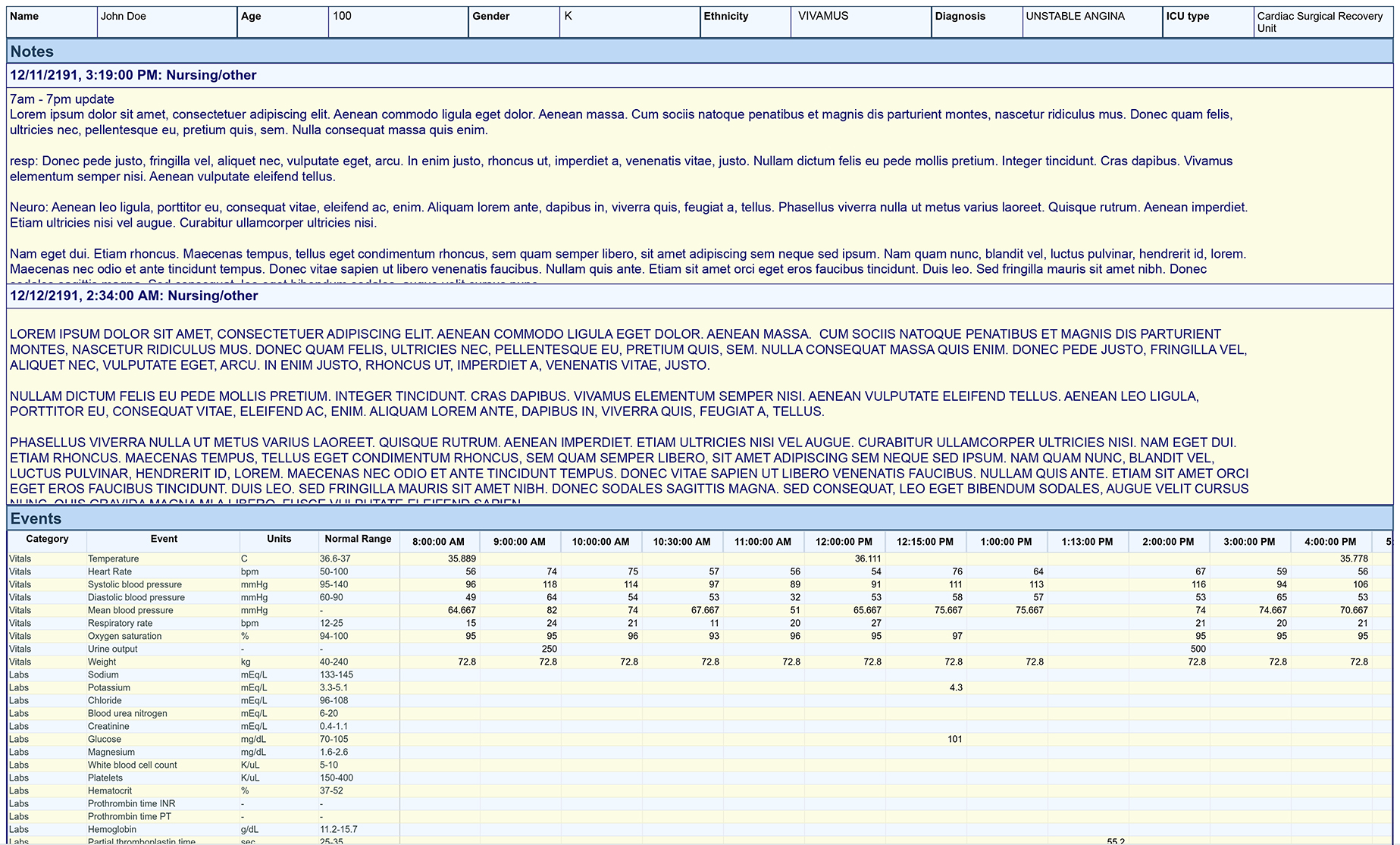}
  \caption{The baseline user interface displaying fake patient data. Note that real patient data was shown to all participants during the study, but cannot be shown here.}
  \label{fig:baselineOverview}
\end{figure*}

\section{Dataset Statistics}
\label{sec:variables} 
From a clinical perspective, the data we present in these interfaces can be described as a mix of categorical variables, numerical variables and textual or free-form data. Nominal categorical variables include demographic information about the patient, admitting diagnosis and ICU type. Additional ordered and unordered categorical data are also contained within free-form nursing and radiology notes. Numerical data are often seen in labs and vitals, in which they are presented either on an interval scale or a ratio scale. All observed variables have temporal attributes associated with them, either per observation (vitals and labs) or for a set of observations (as is in the case of nursing and radiology notes).

\begin{table}[h!]
\centering
\begin{tabular}{l|l}
\hline
Static Variables & Gender\\
& Age \\
& Ethnicity \\
& ICU \\
& Admission Type\\
\hline
 Vitals and Labs & Anion gap \\
 & Bicarbonate \\
 & Blood pH\\
 & Blood urea nitrogen\\
 & Chloride\\
 & Creatinine\\
 & Diastolic blood pressure\\
 & Fraction inspired oxygen\\
 & Glascow coma scale total\\
 & Glucose\\
 & Heart rate\\
 & Hematocrit\\
 & Hemoglobin\\
 & INR\textsuperscript{*}\\
 & Lactate\\
 & Magnesium\\
 & Mean blood pressure\\
 & Oxygen saturation\\
 & Partial thromboplastin time\\
 & Phosphate\\
 & Platelets\\
 & Potassium\\
 & Prothrombin time\\
 & Respiratory rate\\
 & Sodium\\
 & Systolic blood pressure\\
 & Temperature\\
 & Weight\\
 & White blood cell count\\
 & Phosphorus \\
\hline
\multicolumn{2}{l}{\textsuperscript{*}\footnotesize{International normalized ratio of the prothrombin time}}
\end{tabular}
\caption{Variables included in ClinicalVis.}
\label{tab:variables}
\end{table}
% \footnote{A crystalloid bolus is defined as giving a patient a crystalloid volume $ge$ 250 mL in 1 hour. A colloid bolus is a bit less well defined, but we approximate it with $\ge$ 100 mL in 1 hour. Colloid boluses are no longer used, but we include them as they do appear in the data.}

\begin{table}[h!]
  \centering
  \begin{tabular}{|l|l|l|l|l|}
  \hline
  Labels & ID & Gender & Age & ICU \\
  \hline
  \hline
  Control & 13212 & M & 59 & CCU \\
  \cline{2-5}
  & 14474 & M & 74 & MICU \\
  \cline{2-5}
  & 14593 & M & 77 & CCU \\
  \cline{2-5}
  & 5268 & F &56\ & SICU \\
  \cline{2-5}
  & 59381 & F & 45 & SICU \\
  \cline{2-5}
  & 69857 & M & 84 & MICU \\
  \cline{2-5}
  & 9130 & F & 60 & MICU \\
  \cline{2-5}
  & 91513 & F & 57 & SICU \\
  \hline
  VP+ & 32099 & F & 37 & TSICU \\
  \cline{2-5}
  & 7479 & M & 61 & MICU \\
  \hline
  VE+ & 28940 & F & 61 & MICU \\
  \cline{2-5}
  & 48038 & F & 33 & SICU \\
  \hline
  VP+ and VE+ & 1115 & M & 73 & CSRU \\
  \cline{2-5}
  & 14495 & M & 54 & CCU \\
  \cline{2-5}
  & 21454 & F & 70 & CCU \\
  \cline{2-5}
  & 5285 & F & 54 & CSRU \\
  \hline
  Think Aloud & 25328 & F & 78 & CCU \\
  \hline
  
  \end{tabular}
  
  \caption{Selected Patients for the task. ID refers to the subject's MIMIC-III subject identifier. The gender, age and admitting ICU are also reported. Abbreviations: MICU, medical care unit; SICU, surgical care unit; TSICU, trauma surgical care unit; CCU, cardiac care unit; CSRU, cardiac-surgery recovery unit.}
  \label{table:SelectedPatients}
\end{table}

\begin{table}[h!]
  \centering
  \begin{tabular}{|l|l|l|l|l|}
  \hline
  \ & Control & VP+ & VE+ & Both \\
  \hline
  \hline
  Female & 76 & 1 & 9 & 23 \\
  \hline
  Male & 137 & 2 & 12 & 52 \\
  \hline
  \hline
  CCU & 76 & 1 & 5 & 38 \\
  \hline
  CSRU & 29 & 0 & 4 & 20 \\
  \hline
  MICU & 45 & 1 & 6 & 5 \\
  \hline
  NICU & 1 & 0 & 0 & 0 \\
  \hline
  SICU & 41 & 0 & 4 & 5 \\
  \hline
  TSICU & 21 & 1 & 2 & 7 \\
  \hline
  \hline
  Average Age & 72 & 47 & 63 & 68 \\
  \hline
  \end{tabular}
  
  \caption{Demographics of all eligible patients for the task. VP+ refers to patients who received vasopressors. VE+ refers to patients who received invasive ventilation. Abbreviations: MICU, medical care unit; SICU, surgical care unit; TSICU, trauma surgical care unit; CCU, cardiac care unit; CSRU, cardiac-surgery recovery unit.}
  \label{table:SelectedPatients}
\end{table}

\section{Study Prompts}
\label{sec:ThinkQuestions}
Thinkaloud Prompt:
\begin{quotation}
Please walk me through what you see on the screen, and verbalize any thoughts that you have as you arrive at a decision for this patient.
\end{quotation}
Assistive Questions:
\begin{itemize}
\item   What you believe is happening here?
\item   Are you looking for something specific?
\item   What are you looking for?
\item   What action are you trying to perform?
\item   Why you are trying to perform this action? 
\item   What do you expect will happen?
\end{itemize}

\end{document}